\documentclass{IEEEtran}

\usepackage{enumitem,kantlipsum}
\usepackage{graphicx}
\usepackage{epstopdf}
\usepackage{amssymb}
\usepackage{amsmath}
\usepackage{subfigure}
\usepackage{epsfig}
\usepackage{color}
\usepackage{enumitem}
\usepackage{float}
\usepackage{soul}
\usepackage{wrapfig}
\usepackage{arydshln}
\usepackage{tablefootnote}

\usepackage{amsthm}

\def\0{{\bf 0}}
\def\1{{\bf 1}}

\def\beq{\begin{equation*}}
    \def\eeq{\end{equation*}}
\def\bql{\begin{equation}}
    \def\eql{\end{equation}}
\def\bqn{\begin{eqnarray*}}
    \def\eqn{\end{eqnarray*}}
\def\bnl{\begin{eqnarray}}
    \def\enl{\end{eqnarray}}
\def\bma{\begin{bmatrix}}
    \def\ema{\end{bmatrix}}
\def\bmx{\begin{matrix}}
    \def\emx{\end{matrix}}
\def\ben{\begin{enumerate}}
    \def\een{\end{enumerate}}
\def\bit{\begin{itemize}}
    \def\eit{\end{itemize}}
\def\bei{\begin{itemize}}
    \def\eei{\end{itemize}}
\def\bet{\begin{tabular}}
    \def\eet{\end{tabular}}

\newcommand{\ba}{\mathbf{a}}

\newcommand{\be}{\mathbf{e}}

\newcommand{\g}{\mathbf{g}}
\newcommand{\bfs}{\boldsymbol}

\newcommand{\bu}{\mathbf{u}}

\newcommand{\bv}{\mathbf{v}}

\usepackage[dvipsnames]{xcolor}

\def\1{{\bf1}}

\def\b{{\beta}}
\def\a{\alpha}
\def\g{\gamma}

\def\bit{\begin{itemize}}
\def\eit{\end{itemize}}

\def\be{\begin{equation}}
\def\ee{\end{equation}}
\def\ba{\begin{eqnarray}}
\def\ea{\end{eqnarray}}

\def\bes{\begin{equation*}}
\def\ees{\end{equation*}}
\def\bas{\begin{eqnarray*}}
\def\eas{\end{eqnarray*}}
\def\as{{\it a.s.}}

\usepackage{url}
\usepackage{verbatim}

\newtheorem{Remark 1}{Remark}
\newtheorem{Remark 2}[Remark 1]{Remark}
\newtheorem{Remark 3}[Remark 1]{Remark}
\newtheorem{Remark 4}[Remark 1]{Remark}
\newtheorem{Remark 5}[Remark 1]{Remark}
\newtheorem{Remark 6}[Remark 1]{Remark}
\newtheorem{Remark 7}[Remark 1]{Remark}

\newtheorem{Lemma 1}{Lemma}
\newtheorem{Lemma 2}[Lemma 1]{Lemma}
\newtheorem{Lemma 3}[Lemma 1]{Lemma}
\newtheorem{Lemma 4}[Lemma 1]{Lemma}
\newtheorem{Lemma 5}[Lemma 1]{Lemma}
\newtheorem{Lemma 6}[Lemma 1]{Lemma}
\newtheorem{Lemma 7}[Lemma 1]{Lemma}

\newtheorem{Assumption 1}{Assumption}
\newtheorem{Assumption 2}[Assumption 1]{Assumption}
\newtheorem{Assumption 3}[Assumption 1]{Assumption}
\newtheorem{Assumption 4}[Assumption 1]{Assumption}
\newtheorem{Definition 1}{Definition}
\newtheorem{Theorem 1}{Theorem}
\newtheorem{Theorem 2}[Theorem 1]{Theorem}
\newtheorem{Theorem 3}[Theorem 1]{Theorem}
\newtheorem{Theorem 4}[Theorem 1]{Theorem}
\newtheorem{Theorem 5}[Theorem 1]{Theorem}
\newtheorem{Theorem 6}[Theorem 1]{Theorem}
\newtheorem{Theorem 7}[Theorem 1]{Theorem}
\newtheorem{Theorem 8}[Theorem 1]{Theorem}
\newtheorem{Theorem 9}[Theorem 1]{Theorem}
\newtheorem{Theorem 10}[Theorem 1]{Theorem}

\newtheorem{Proposition 1}{Proposition}
\newtheorem{Proposition 2}[Proposition 1]{Proposition}

\title{\LARGE \bf
Ensuring both { Provable} Convergence and  Differential Privacy in Nash Equilibrium Seeking on Directed Graphs}

\author{Yongqiang Wang, Tamer Ba{\c{s}}ar
\thanks{Yongqiang Wang is with the Department of Electrical and Computer Engineering, Clemson University, Clemson, SC 29634, USA
{\tt\small{yongqiw}@clemson.edu}
}%
\thanks{Tamer Ba{\c{s}}ar is with the Coordinated Science Lab, University of Illinois
at Urbana-Champaign, Urbana, IL 61801, USA {\tt\small
basar1@illinois.edu}}
  }

\begin{document}

\maketitle
\thispagestyle{empty}
\pagestyle{empty}

\begin{abstract}
We propose a distributed Nash equilibrium seeking approach that can achieve both { provable} convergence and rigorous differential privacy with finite cumulative privacy budget, which is in sharp contrast to existing differential-privacy solutions for networked games that have to trade  provable convergence for differential privacy. The approach is applicable   when the communication graph is directed and unbalanced. 
Numerical comparison results with existing counterparts confirm the effectiveness of the proposed approach.
\end{abstract}

\section{Introduction}
Nash equilibrium  (NE) seeking in game theory addresses the problem where multiple players compete to minimize their individual cost functions \cite{frihauf2011nash}.
 In many  application scenarios,   individual players  only have access to the decisions of their local neighbors, which is usually termed as games in the partial-decision information setting \cite{pavel2019distributed,belgioioso2019distributed}. In contrast to the classical full-decision information setting where a player knows the past actions of all other players, in  the partial-decision information setting, individual players cannot evaluate their  cost functions or gradients due to  lack of necessary  information. Consequently, players have to exchange action information with their local neighbors   for NE seeking.

  Significant inroads have been made in fully distributed NE seeking (see, e.g., \cite{koshal2016distributed,salehisadaghiani2016distributed,tatarenko2020geometric,bianchi2022fast,nguyen2022distributed}). However, all of these distributed algorithms require players to share explicit (estimated) decisions in every iteration, which is problematic when sensitive information is involved. In fact, given that in noncooperative games the players are not fully cooperative, it is important for individual players to protect their private information, which, otherwise, might be exploited by others.
Recently, several results have been reported on privacy protection in NE seeking (see, e.g., \cite{lu2015game,shilov2021privacy}). However, most of these results assume the presence of a coordinator. In the fully distributed case, the authors in \cite{gade2020privatizing} exploit  spatially-correlated noise to protect the privacy of players. However, their approach is only effective when the communication graph satisfies certain properties. Recently, the authors of \cite{shakarami2022distributed} have used a constant uncertain parameter to obfuscate individual players' pseudo-gradients  to achieve privacy protection in continuous-time aggregative games. However, the privacy  strength enabled by such a constant scalar is weak in the sense that only the exact value of the cost function is prevented  from being uniquely identifiable.  As differential privacy has emerged as the de facto standard for privacy protection due to its  strong resilience against arbitrary post-processing and  auxiliary information \cite{dwork2014algorithmic},  recent results in \cite{ye2021differentially} and \cite{wang2022differentially} propose  differential-privacy mechanisms for   aggregative games, which, however, have to sacrifice { provable} convergence to the exact Nash equilibrium.

In this paper, we introduce  a distributed NE seeking approach on directed graphs that can ensure both { provable} convergence and rigorous $\epsilon$-differential privacy with guaranteed finite cumulative privacy budget. 
We propose to gradually weaken the inter-player interaction to attenuate the effect of differential-privacy noise in shared messages on NE seeking. Note that inter-player interaction is necessary for all players' convergence to the NE (which exists and is unique in our case), and thus we  judiciously design  the weakening factor sequence and the stepsize sequence, under which we prove that our approach can ensure provable convergence to the exact unique  NE even in the presence of differential-privacy noise. We   prove that the algorithm is $\epsilon$-differentially private with a finite cumulative privacy budget, even when the number of iterations tends to infinity.  It is worth noting that compared with our recent results on differentially-private distributed optimization \cite{wang2022tailoring,wang2022quantization}, the results here for NE seeking are fundamentally different: agents in distributed optimization are cooperative in computing a common objective function, whereas players in games are competitive and only mind their own individual cost functions. Moreover, different from our recent results in \cite{wang2022aggregate} which address aggregative games on symmetric communication graphs, this paper addresses general networked games that are not necessarily aggregative  on directed communication graphs that could be unbalanced.

{\bf Notations:}  We use $\mathbb{R}^d$ to denote the Euclidean space of
dimension $d$. We write $I_d$ for the identity matrix of dimension $d$,
and ${\bf 1}_d$ for  the $d$-dimensional  column vector with all
entries equal to 1. 
For a
vector $x$, $[x]_i$ denotes its $i$th element.
We write $x> 0$ (resp. $x\geq 0$) if all elements of
$x$ are positive (resp. non-negative).
  We use $\langle\cdot,\cdot\rangle$ to denote the inner product and
 $\|x\|_2$ for the standard Euclidean norm of a vector $x$. We use $\|x\|_1$ to represent the $\ell_1$ norm of a vector $x$.
We write $\|A\|$ for the matrix norm induced by a vector norm $\|\cdot\|$.
 For two  vectors  $u$ and $v$ with the same dimension, we write $u\leq v$ to mean that each entry of $u$ is no larger than the corresponding entry of $v$. { We use {\it a.s.} to denote {\it almost surely} or {\it almost sure}, depending on the syntax context}.


\section{Problem Formulation and Preliminaries}\label{sec-problem}

\subsection{On Networked Nash Games}
We consider  a networked Nash game among a set of $m$ players, i.e., $[m]=\{1,2,\ldots,m\}$. Player $i$ is characterized by a feasible action set $K_i \subseteq\mathbb{R}^{d_i}$ and a cost function $f_i(x_i,x_{-i})$ where $x_i\in K_i$ is the decision of player $i$ and $x_{-i}\triangleq[x_1^T,\cdots,x_{i-1}^T,x_{i+1}^T,\cdots,x_m^T]^T$
 denotes the joint decisions of all players except  player $i$. Note that we allow different $x_i$ to have different dimensions $d_i$.

Traditionally when a mediator/coordinator exists, every player $i$ can access all other players' decisions $x_{-i}$. Then, the game that player
$i$ faces can be formulated as:
\begin{equation}\label{eq:formulation}
 \min f_i(x_i,x_{-i})\quad {\rm s.t.}\quad x_i\in K_i\:\: {\rm and}\:\: x_{-i}\in K_{-i}.
\end{equation}
The   function $f_i(\cdot)$ is assumed to be known to player $i$ only.

At the NE  $ x^{\ast}=
[(x_1^{\ast})^T, \ldots,(x_m^{\ast})^T]^T\in\mathbb{R}^{D}$ with $D=\sum_{i=1}^{m}d_i$, each player has
$
f_i(x_i^\ast,x_{-i}^\ast)\leq f_i(x_i,x_{-i}^\ast), \forall x_i\in K_i.
$
Namely, at the NE, no player can unilaterally reduce its cost by changing its own decision.

We consider a  scenario where no mediator/coordinator exists, and    players  share decisions locally among neighbors, which is commonly referred to as the partial-decision information scenario \cite{pavel2019distributed}.  We use a directed graph $\mathcal{G}=([m],\mathcal{E})$ to denote the  communication pattern where $[m]=\{1,2,\ldots,m\}$ is the set of nodes (players) and $\mathcal{E}\subseteq [m]\times [m]$  is the edge set of ordered node pairs describing the interactions among players. We also use the notion of directed graph induced by a  weight  matrix $L=\{L_{ij}\}\in\mathbb{R}^{m\times m}$, denoted as $\mathcal{G}_L=([m],\mathcal{E}_L)$. More specifically, in $\mathcal{G}_L=([m],\mathcal{E}_L)$, a directed edge $(i,j)$ from agent $j$ to agent $i$ exists,
 i.e., $(i,j)\in \mathcal{E}_L$ if and only if $L_{ij}>0$.
For a player $i\in[m]$,
its in-neighbor set
$\mathbb{N}^{\rm in}_i$ is defined as the collection of players $j$ such that $L_{ij}>0$; similarly,
the out-neighbor set $\mathbb{N}^{\rm out}_i$ of player $i$ is the collection of players $j$ such that $L_{ji}>0$. 

\begin{Assumption 2}\label{as:monotone}
 $K_i =\mathbb{R}^d$ and every $f_i(x_i,x_{-i})$ is convex and  differentiable in $x_{i}$ over $\mathbb{R}^d$ for each $x_{-i}$.
\end{Assumption 2}

To characterize  the NE of the  game (\ref{eq:formulation}),   we also define
\vspace{-0.15cm}
\begin{equation}\label{eq:F_i}
  \phi(x)\triangleq  \left(  F_1^T(x_1,x_{-1}), \cdots,F^T_m(x_m,x_{-m})  \right)^T,
\end{equation}
where $F_i(x_i,x_{-i})\triangleq \nabla_{x_i} f_i(x_i,x_{-i}).$

\begin{Assumption 2}\label{as:Lipschitz}
 $\phi(x)$ is { strongly} monotone over $K\triangleq K_1\times\cdots\times K_m$, i.e., for all $x\neq x'$ in $K$,
$
\left(\phi(x)-\phi(x')\right)^T(x-x')\geq\mu\|x-x'\|
$ holds for some $\mu>0$.  Each mapping $F_i(x_i,x_{-i})$ is Lipschitz continuous in   both of its arguments, $x_i$ and $x_{-i}$. Namely, for all $x_i,y_i\in\mathbb{R}^{d_i}$ and $x_{-i},y_{-i}\in\mathbb{R}^{D-d_i}$,  where $D=\sum_{i=1}^{m}d_i$, 
$
\|F_i(x_i,x_{-i})-F_i(y_i,x_{-i}) \|_2\leq { K_1}\|x_{i}-y_{i}\|_2$ and
$\|F_i(x_i,x_{-i})-F_i(x_i,y_{-i}) \|_2\leq { K_2}\|x_{-i}-y_{-i}\|_2
$
hold for all $i\in[m]$, where ${ K_1}$ , ${ K_2}$ are some  constants.
\end{Assumption 2}


  Assumption \ref{as:Lipschitz} ensures that  (\ref{eq:formulation}) has a unique NE $ x^{\ast}$ \cite{scutari2014real}.


\begin{Assumption 1}\label{Assumption:coupling}
 The off-diagonal entries of the matrix  $L=\{L_{ij}\}\in \mathbb{R}^{m\times m}$ are non-negative  and its diagonal entries  $L_{ii}=-\sum_{j=1}^{m}L_{ij}$ satisfy $L_{ii}>-1$ for all $i\in[m]$. Moreover, the digraph $\mathcal{G}_L$ is strongly connected.
\end{Assumption 1}

 In the analysis of our approach, we use the following results:
\begin{Lemma 2}({ Lemma 2 of \cite{wang2022tailoring}})\label{Lemma-polyak_2}
Let $\{v^k\}$,$\{\a^k\}$, and $\{p^k\}$ be random nonnegative  scalar sequences, and
$\{q^k\}$ be a deterministic nonnegative  scalar sequence satisfying
$\sum_{k=0}^\infty \a^k<\infty$  \as,
$\sum_{k=0}^\infty q^k=\infty$, $\sum_{k=0}^\infty p^k<\infty$ \as,
and
\[
\mathbb{E}\left[v^{k+1}|\mathcal{F}^k\right]\le(1+\a^k-q^k) v^k +p^k,\quad \forall k\geq 0\quad\as
\]
where $\mathcal{F}^k=\{v^\ell,\a^\ell,p^\ell; 0\le \ell\le k\}$.
Then, $\sum_{k=0}^{\infty}q^k v^k<\infty$ and
$\lim_{k\to\infty} v^k=0$ hold almost surely.
\end{Lemma 2}
\begin{Lemma 1}({ Lemma 5 of \cite{wang2022tailoring}})\label{le:vector_convergence}
Let  $\{\bv^k\}\subset \mathbb{R}^d$
and $\{\bu^k\}\subset \mathbb{R}^p$ be random nonnegative
vector sequences, and $\{a^k\}$ and $\{b^k\}$ be random nonnegative scalar sequences   such that
\[
\mathbb{E}\left[\bv^{k+1}|\mathcal{F}^k\right]\le (V^k+a^k{\bf 1}{\bf1}^T)\bv^k +b^k{\bf 1} -H^k\bu^k,\quad \forall k\geq 0
\]
holds \as, where $\{V^k\}$ and $\{H^k\}$ are random sequences of
nonnegative matrices and
$\mathbb{E}\left[\bv^{k+1}|\mathcal{F}^k \right]$ denotes the conditional expectation given
 $\bv^\ell,\bu^\ell,a^\ell,b^\ell,V^\ell,H^\ell$ for $\ell=0,1,\ldots,k$.
Assume that $\{a^k\}$ and $\{b^k\}$ satisfy
$\sum_{k=0}^\infty a^k<\infty$ and $\sum_{k=0}^\infty b^k<\infty$ a.s., and
that there exists a (deterministic) vector $\pi>0$ such that
$\pi^T V^k\le \pi^T$ and $\pi^TH^k\ge 0$ hold a.s. for all $k\geq 0$.
Then, we have
1) $\{\pi^T\bv^k\}$ converges to some random variable $\pi^T\bv\geq 0$ a.s.; 2) $\{\bv^k\}$ is bounded a.s.; and
3) $\sum_{ k=0 }^\infty \pi^TH^k\bu^k<\infty$ holds almost surely.
\end{Lemma 1}

\subsection{On Differential Privacy}
We adopt the notion of $\epsilon$-differential privacy (DP) for continuous bit streams \cite{dwork2010differential}, which has recently been applied to distributed optimization  (see, e.g., \cite{Huang15} as well as our work \cite{wang2022tailoring}). To enable DP, we  inject  Laplace   noise ${\rm Lap}(\nu)$ to all shared messages, where  $\nu>0$ is a constant parameter of   the probability density function $\frac{1}{2\nu}e^{-\frac{|x|}{\nu}}$. One can verify that ${\rm Lap}(\nu)$ has mean zero and variance $2\nu^2$.
We represent the networked game $\mathcal{P}$ in (\ref{eq:formulation}) by three parameters ($K, \mathbb{F},\mathcal{G}_L$), where
 $K\triangleq K_1\times \cdots\times K_m$    is the domain of decision variables,   $ \mathbb{F} \triangleq\{ f_1,\,\cdots,f_m\}$, and $\mathcal{G}_L$   denotes the communication graph. We define ``adjacency" between two  games as follows:

\begin{Definition 1}\label{de:adjacency}
Two networked games $\mathcal{P}\triangleq (K, \mathbb{F},\mathcal{G}_L)$ and $\mathcal{P}'\triangleq (K', \mathbb{F}',\mathcal{G'}_L)$ are adjacent if the following conditions hold:
\begin{itemize}
\item $K=K'$   and $\mathcal{G}_L=\mathcal{G}_L'$, i.e., the domain of decision variables and the communication graphs are identical;
\item there exists an $i\in[m]$ such that $f_i\neq f_i'$ but $f_j=f_j'$ for all $j\in[m],\,j\neq i$.
\end{itemize}
\end{Definition 1}


 Given a distributed algorithm for NE seeking, we represent an execution of this algorithm as $\mathcal{A}$, which is an infinite sequence of the iteration variable $\vartheta$, i.e., $\mathcal{A}=\{\vartheta^0,\vartheta^1,\cdots\}$. We consider adversaries that can observe all communicated messages in the network $\mathcal{G}_L$. Thus, the observation part of an execution is the infinite sequence of shared messages, which is represented by $\mathcal{O}$. We define the mapping from execution sequence to observation sequence under a networked game $\mathcal{P}$ by $\mathcal{R}_{\mathcal{P},{\vartheta}^0}(\mathcal{A})\triangleq \mathcal{O}$, where ${\vartheta}^0$ denotes the initial condition. Given a networked game $\mathcal{P}$, an initial condition ${\vartheta}^0$,  and observation sequence $\mathcal{O}$, 
   { $\mathcal{R}_{\mathcal{P},{\vartheta}^0}^{-1}(\mathcal{O})$} is the set of executions $\mathcal{A}$ that can generate the observation $\mathcal{O}$.
 \begin{Definition 1}
   ($\epsilon$-DP \cite{Huang15}). For a given $\epsilon>0$, an  NE seeking algorithm is $\epsilon$-differentially private if for any two adjacent { networked games} $\mathcal{P}$ and $\mathcal{P}'$, any set of observation sequences $\mathcal{O}_s\subseteq\mathbb{O}$ ($\mathbb{O}$ denoting the set of all possible observation sequences), and any initial state ${\vartheta}^0$, we always have
$
        \mathbb{P}[\mathcal{R}_{\mathcal{P}, {\vartheta}^0}^{-1}\left(\mathcal{O}_s\right)]\leq e^\epsilon\mathbb{P}[\mathcal{R}_{\mathcal{P}', {\vartheta}^0}^{-1}\left( \mathcal{O}_s, \right)],
 $
    where the probability $\mathbb{P}$ is taken over the randomness over iteration processes.
 \end{Definition 1}

The definition of $\epsilon$-DP guarantees that an adversary having access to all shared messages  cannot gain information with a  significant probability of any  player's cost function. It also implies   a smaller $\epsilon$ corresponding to  a higher level of privacy. 

\section{A differentially-private NE seeking algorithm}\label{se:algorithm1}

We present in this section a fully distributed NE seeking algorithm (Algorithm 1 below) which can guarantee both $\epsilon$-DP and exact convergence.

\noindent\rule{0.49\textwidth}{0.5pt}
\noindent\textbf{Algorithm 1: Distributed NE seeking with provable convergence and differential privacy}
\noindent\rule{0.49\textwidth}{0.5pt}

\begin{enumerate}[wide, labelwidth=!, labelindent=0pt]
    \item[] Parameters: Stepsize $\lambda^k>0$ and
    weakening factor $\gamma^k>0$.
    \item[] Every player $i$ maintains one decision variable  $x_{(i)i}^k$, and $m-1$ estimates $x_{(i)-i}^k\triangleq [(x_{(i)1}^k)^T,\,\cdots,\,(x_{(i)i-1}^k)^T,\,(x_{(i)i+1}^k)^T,\,\cdots,(x_{(i)m}^k)^T]^T $ of other players' decision variables. Player $i$ sets $x^0_{(i)\ell}$  randomly  in $\mathbb{R}^{d_\ell}$ for all $\ell\in[m]$.
    \item[] {\bf for  $k=1,2,\ldots$ do}
        \begin{enumerate}
        \item For both its decision variable $x_{(j)j}^k$ and estimate variables
    $x_{(j)1}^k,\,\cdots,\,x_{(j)j-1}^k,\,x_{(j)j+1}^k,\,\cdots,x_{(j)m}^k $, every player $j$ adds respective persistent DP noise   $\zeta_{(j)1}^{k},\,\cdots,\,\zeta_{(j)m}^{k}$,  and then sends the obscured values $x_{(j)1}^k+\zeta_{(j)1}^{k},\cdots,\,x_{(j)m}^k+\zeta_{(j)m}^{k}$ to all players
        $i\in\mathbb{N}_j^{\rm out}$.
        \item After receiving  $x_{(j)1}^k+\zeta_{(j)1}^{k},\cdots,\,x_{(j)m}^k+\zeta_{(j)m}^{k}$ from all $j\in\mathbb{N}_i^{\rm in}$, player $i$ updates its decision and estimate variables:
        \begin{equation}\label{eq:update_in_Algorithm1}
        \begin{aligned}
           \hspace{-0.4cm}  x_{(i)i}^{k+1}\hspace{-0.1cm}=& x_{(i)i}^k \hspace{-0.05cm}+\hspace{-0.05cm}\gamma^k\hspace{-0.2cm}\sum_{j\in \mathbb{N}_i^{\rm in}}\hspace{-0.2cm} L_{ij}(x_{(j)i}^k \hspace{-0.05cm}+\hspace{-0.05cm}\zeta_{(j)i}^k\hspace{-0.08cm}-\hspace{-0.08cm}x_{(i)i}^k) \hspace{-0.08cm}-\hspace{-0.08cm}\lambda^k\hspace{-0.05cm} F_i\hspace{-0.03cm}(\hspace{-0.05cm}x_{(i)i}^k,\hspace{-0.06cm}x_{(i)-i}^k\hspace{-0.03cm}),\\
            \hspace{-0.4cm} x_{(i)\ell}^{k+1}=&x_{(i)\ell}^k\hspace{-0.05cm}+\hspace{-0.05cm}\gamma^k\hspace{-0.15cm}\sum_{j\in \mathbb{N}_i^{\rm in}} \hspace{-0.2cm}L_{ij}(x_{(j)\ell}^k+\zeta_{(j)\ell}^k-x_{(i)\ell}^k),\:\: \forall \ell\neq i.
        \end{aligned}
        \end{equation}
        \item {\bf end}
        \vspace{-0.2cm}
    \end{enumerate}
\end{enumerate}
\vspace{-0.1cm} \rule{0.49\textwidth}{0.5pt}
{
\vspace{-0.5cm}
\begin{Remark 1}
The sequences   $\{\gamma^k\}$ and $\{\lambda^k\}$, and the DP noise parameter   are hard-coded into all players' programs and  need no adjustment/coordination in implementation.
\end{Remark 1}
}
\vspace{-0.2cm}
\section{A General Convergence Result}\label{sec:general_convergence}

We first have to establish some general results necessary for the convergence analysis of Algorithm 1.
\begin{Lemma 1}\label{lemma:coupling}
Under Assumption \ref{Assumption:coupling}, we have the following properties { when $\gamma^k>0$ in Algorithm 1 is sufficiently small}:
\begin{enumerate}
  \item the eigenvectors of  the matrix $I+\gamma^kL$ are time-invariant;
  \item $I+\gamma^kL$ has a unique
  positive left eigenvector $u^T$ (associated with eigenvalue 1) satisfying $u^T{\bf 1}=m$;
  \item the  spectral radius of   $I+\gamma^kL-\frac{{\bf 1}u^T}{m}$ is upper-bounded  by $1-\alpha\gamma^k$, where $0<\alpha<1$.
  \item there exists an $L$-dependent matrix norm $\|\cdot\|_L$ such that $\|I+\gamma^kL-\frac{{\bf 1}u^T}{m}\|_L\leq 1-\alpha\gamma^k$ for $0<\alpha<1$ when $\gamma^k$ is small enough. Moreover, this norm has an associated inner product $\langle\cdot,\cdot\rangle_L$, i.e.,  $\|x\|_L^2=\langle x,x\rangle_L$.
\end{enumerate}

\end{Lemma 1}
\begin{proof}
\begin{enumerate}[wide, labelwidth=!,labelindent=8pt]
  \item Representing the eigenvalues and associated eigenvectors of $L$ as $\{\varrho_1,\,\cdots,\varrho_m\}$ and $\{v_1,\,\cdots,v_m\}$, respectively, we can verify that the eigenvalues and associated eigenvectors of $I+\gamma^kL$ are given by $\{1+\gamma^k\varrho_1,\,\cdots,1+\gamma^k\varrho_m\}$ and $\{v_1,\,\cdots,v_m\}$, respectively. 
  \item One can obtain from  \cite{horn2012matrix} (or Lemma 1 in \cite{pu2020push}) that $I+L$ has a unique   positive  left eigenvector $u^T$ (associated with eigenvalue 1) satisfying $u^T{\bf 1}=m$. Hence, using  statement 1),   $I+\gamma^kL$ has a unique   positive  left eigenvector $u^T$ (associated with eigenvalue 1) satisfying $u^T{\bf 1}=m$.
       \vspace{0.1cm}
  \item Representing the eigenvalues of $L$ by $\{\varrho_1,\,\cdots,\,\varrho_m\}$, the eigenvalues of $I+L$ can be expressed as $\{1+\varrho_1,\,\cdots,\,1+\varrho_m\}$. Under Assumption \ref{Assumption:coupling}, $I+L$ is irreducible. Using the Perron–Frobenius theorem, one can obtain  that  $I+L$ has one unique eigenvalue equal to one and all its other eigenvalues  strictly less than one in absolute value, implying that one and only one of $\varrho_i$ is zero. Represent this eigenvalue of $L$ as $\varrho_m=0$ without loss of generality. Then we have  $|1+\varrho_i|<1$ for all $1\leq i\leq m-1$. One can verify that the eigenvalues of $I+\gamma^kL-\frac{{\bf 1}u^T}{m}$ are given by $\{1+\gamma^k\varrho_1,\,\cdots,1+\gamma^k\varrho_{m-1},\,0\}$. Next, we prove the third statement by showing that there exists an $\alpha$ satisfying
      $|1+\gamma^k\varrho_{i}|<1-\alpha\gamma^k$ for every $i=1,2,\cdots,m-1$.

       We  represent $\varrho_i$ as $\varrho_i=a_i+{\rm i}b_i$, where $a_i$ and $b_i$ are real numbers, and $\rm i$ is the imaginary unit. Because $|1+\varrho_i|<1$ holds for all $i=1,2,\cdots,m-1$, we have $a_i<0$ for $i=1,2,\cdots,m-1$. Under the new representation of $\varrho_i$, $|1+\gamma_k\varrho_i|$ becomes $\sqrt{(1-|a_i|\gamma^k)^2+(b_i\gamma^k)^2}$. So we only have to prove
   $
          \sqrt{(1-|a_i|\gamma^k)^2+(b_i\gamma^k)^2}<1-\gamma^k\alpha
    $
         for some $0<\alpha<1$ when $\gamma^k$ is small enough.
        { Squaring  both sides of the inequality yields} $
        \alpha^2(\gamma^k)^2-2\alpha\gamma^k>(\gamma^k)^2(a_i^2+b_i^2)-2\gamma^k|a_i|,
        $
        i.e., \vspace{-0.1cm}
        \begin{equation}\label{eq:spectral_2}
        \alpha^2 -\textstyle\frac{2}{\gamma^k}\alpha> (a_i^2+b_i^2)-\frac{2 |a_i|}{\gamma^k}.
        \vspace{-0.15cm}
        \end{equation}
       When $\gamma^k$ is less than $\frac{2|a_i|}{a_i^2+b_i^2}$ (note $a_i<0$ for $i=1,\cdots,m-1$), the right hand side of (\ref{eq:spectral_2}) is negative whereas the left hand side is a quadratic function of $\alpha$ with two   x-intercepts given by $\alpha=0$ and $\alpha=\frac{2}{\gamma^k}$. So there always exits an $\alpha$ in the interval $(0,\,1)$ making the left hand side of (\ref{eq:spectral_2}) larger than its right hand side, and hence making (\ref{eq:spectral_2}) hold. Therefore, there always exists an $0<\alpha<1$ making $|1+\gamma_k\varrho_i|<1-\alpha\gamma^k$ hold when $\gamma^k>0$ is less than $\frac{2|a_i|}{a_i^2+b_i^2}$.

       Given that the above derivation is independent of $i$, we have   $|1+\gamma_k\varrho_i|<1-\alpha\gamma^k$  for some $0<\alpha<1$ and all $i=1,\cdots,m-1$, and hence the third statement in the Lemma.
     \vspace{0.1cm}
       \item
       { According to   \cite{wang2023spectral_norm}, there exits a matrix norm $\|\cdot\|_L$,  which is only dependent on the eigenvectors (or unitary  Schur-decomposition matrix in the more general complex-matrix case) of $I+\gamma^kL-\frac{{\bf 1}u^T}{m}$, such that the norm of $I+\gamma^kL-\frac{{\bf 1}u^T}{m}$ is arbitrarily close to its spectral radius. Condition 2) proves that the eigenvectors of $I+\gamma^kL-\frac{{\bf 1}u^T}{m}$ are time-invariant and independent of $\gamma^k$. Hence, the matrix norm  $\|\cdot\|_L$ is independent of $\gamma^k$. Using Condition 3) yields that $\|I+\gamma^kL-\frac{{\bf 1}u^T}{m}\|_L<1-\alpha\gamma^k$ holds for  some $0<\alpha<1$ when $\gamma^k$ is smaller than $\min\{\frac{2|a_1|}{a_1^2+b_1^2},\,\cdots,\frac{2|a_{m-1}|}{a_{m-1}^2+b_{m-1}^2}\}$. Moreover, from \cite{wang2023spectral_norm}, $\|\cdot\|_L$ can be expressed as $\|x\|_L=\|\hat{L}x\|_2$ for some $\hat{L}$ determined by $L$. Thus, the norm $\|\cdot\|_L$ satisfies the   Parallelogram Law and,
hence, has an associated inner product  $\langle \cdot,\rangle_L$. We refer the reader to  Sec. II.B of \cite{xin2019distributed} for an instantiation of this norm and inner product (which  only depends on the left eigenvector $u^T$) for directed graphs.}
\end{enumerate}
\vspace{-0.2cm}
\end{proof}

Using the time-invariant positive left eigenvector $u\triangleq[u_1,\cdots,u_m]^T$ of $I+\gamma^kL$ from Lemma 3, we define a weighted average  $\bar{x}_i^k\triangleq \frac{1}{m}\sum_{\ell=1}^{m}{ u_{\ell}}x_{(\ell)i}^k$ of player $i$'s decision variable $x_{(i)i}^k$ and other players' estimates $x_{(j)i}^k$ ($j\neq i$) of this decision variable.

For the convenience of analysis, we also define the assembly of the $i$th decision variable ${\bf x}_i^k$ as well as the assembly of the weighted average   $\bar{\bf x}_i^k$ as
\begin{equation}\label{eq:bf_x_i}
{\bf x}_i^k=\left[\begin{array}{c}(x_{(1)i}^k)^T\\ \vdots\\(x_{(m)i}^k)^T\end{array}\right]\in\mathbb{R}^{m\times {d_i}}
,\:
\bar{\bf x}_i^k=\left[\begin{array}{c}(\bar{x}_{i}^k)^T\\ \vdots\\(\bar{x}_{i}^k)^T\end{array}\right]\in\mathbb{R}^{m\times {d_i}}.
\end{equation}

To measure  the distance between matrix variables ${\bf x}_i^k$ and $\bar{\bf x}_i^k$,  we define a matrix norm for an arbitrary  vector norm { $\|\cdot\|_p$}. Specifically, for a matrix $X\in\mathbb{R}^{m\times {d_i}}$,  we  define
{ $
\|X\|_p\triangleq \left\|\left[\|X_{(1)}\|_p,\,\|X_{(2)}\|_p,\cdots,\|X_{(d_i)}\|_p \right] \right\|_2$},
 where   $X_{(i)}$ denotes the $i$th column of $X$. $\|{\bf x}_i^k-{\bf\bar{x}}_i^k \|_L$ measures the distance between all players'  $x_{(j)i}^k$ for $j\in[m]$ and their average $\bar{x}_i^k$.

Based on the inner product $\langle \cdot,\cdot\rangle_L$ for vectors, we also define an inner product for matrices consistent with the $\|\cdot\|_L$ norm for matrices. More specifically, for a matrix $X$ in $\mathbb{R}^{m\times d_i}$,  we  define  $\langle X,X\rangle_L$ as:
$
\langle X,X\rangle_L=\sum_{i=1}^{d_i}\langle X_{(i)},X_{(i)}\rangle_L$,
where $X_{(i)}$ denotes the $i$th column of $X$.
Since for any  column  $X_{(i)}$ of a matrix $X$, we have $\|X_{(i)}\|_L^2=\langle X_{(i)},X_{(i)}\rangle_L$, one can verify that $\|X\|_L^2=\langle X,X\rangle_L$ holds for any matrix $X\in\mathbb{R}^{m\times d_i}$.
In addition, we have the following result:
\begin{Lemma 1}\label{lemma:normR_norm2}
For any norm $\|\cdot\|_p$, $X\in\mathbb{R}^{m \times d_i}$,   and $W\in\mathbb{R}^{m \times m}$,   we always have
$
\|WX\|_p\leq \|W\|_p\|X\|_p
$.
Furthermore, there exist constants { $\delta_{L,2}$} and $\delta_{2,L}$ such that   $\|X\|_L\leq \delta_{L,2}\|X\|_2$ and $\|X\|_2\leq \delta_{2,L}\|X\|_L$ hold for  any $X\in\mathbb{R}^{m\times d_i}$.
\end{Lemma 1}
\begin{proof}
  The proof follows from the line of reasoning in Lemma 5 and Lemma 6 in \cite{pu2020push}, and hence is not included  here.
\end{proof}

Based on the above results, we  have the following convergence result for general distributed algorithms for problem (\ref{eq:formulation}):

\begin{Proposition 1}\label{th-main_decreasing}
Under Assumptions \ref{as:monotone} and \ref{as:Lipschitz}, let   $ x^{\ast}=
[(x_1^{\ast})^T, \ldots,(x_m^{\ast})^T]^T$ denote the unique  NE of (\ref{eq:formulation}). If, under the interaction matrix $L$, { any distributed NE-seeking algorithm} generates sequences
$\{{\bf x}_i^k\}$ for all $i\in[m]$ such that
\as\,  we have
\begin{equation}\label{eq:Theorem_decreasing}
\begin{aligned}
&\left[\begin{array}{c}
\mathbb{E}\left[\sum_{i=1}^m\|\bar{x}_i^{k+1}-x_i^*\|_2^2|\mathcal{F}^k\right]\cr
\mathbb{E}\left[\sum_{i=1}^m\| {\bf x}_i^{k+1} - {\bf \bar{x}}_i^{k+1}  \|_L^2|\mathcal{F}^k\right]\end{array}
\right]
\\
&\le \left( \left[\begin{array}{cc}
1 &  \kappa_1\gamma^k \cr
0& 1-\kappa_2\gamma^k\cr
\end{array}\right]
+a^k {\bf 1}{\bf 1}^T\right)\left[\begin{array}{c}\sum_{i=1}^m\|\bar{x}_i^{k}-x_i^*\|_2^2\cr
\sum_{i=1}^m\|{\bf x}_i^{k} - {\bf \bar{x}}_i^{k} \|_L^2\end{array}\right]&&\cr
&\quad+b^k{\bf 1} - c^k \left[\begin{array}{c}
 \left(\phi(\bar{x}^k)- \phi(x^*)\right)^T(\bar{x}^k-x^{\ast}) \cr
 0\end{array}\right],\quad\forall k\geq 0
 \end{aligned}
\end{equation}
where
{
\begin{itemize}
\item $\|\cdot\|_L$ is an $L$-dependent norm,
$\mathcal{F}^k=\{{\bf x}_i^\ell, \, i\in[m],\, 0\le \ell\le k\}$, $\bar{x}^k=\left[(\bar{x}_1^k)^T,\,\cdots,(\bar{x}_m^k)^T\right]^T$;
\item the random nonnegative scalar sequences $\{a^k\}$, $\{b^k\}$ satisfy $\sum_{k=0}^\infty a^k<\infty$ and $\sum_{k=0}^\infty b^k<\infty$, respectively, a.s., and the deterministic  nonnegative sequences $\{c^k\}$ and $\{\gamma^k\}$  satisfy
$
\sum_{k=0}^\infty c^k=\infty$ and $
\sum_{k=0}^\infty \gamma^k=\infty
$;
\item  the scalars $\kappa_1$ and $\kappa_2$ satisfy  $\kappa_1>0$ and $0<\kappa_2\gamma^k<1$, respectively, for all $k\geq 0$.
\end{itemize}
}
Then,
$\lim_{k\to\infty}\|{\bf x}_i^{k} - {\bf \bar{x}}_i^{k}\|_L=0$
and $\lim_{k\to\infty}\|\bar{x}_i^k-x_i^*\|=0$ a.s.  for all   $i$, implying
$\lim_{k\to\infty}\|x_{(i)i}^k-x_i^*\|=0$ a.s.  for all   $i$.
\end{Proposition 1}

\begin{proof}
Since we have
 $\left(\phi(\bar{x}^k)- \phi(x^*)\right)^T(\bar{x}^k-x^{\ast})>0$ for all $k$ from Assumption \ref{as:monotone},  by letting $\bv^k=\left[\sum_{i=1}^{m}\|\bar{x}_i^k-x_i^*\|_2^2,\ \sum_{i=1}^m \|{\bf x}_i^{k} - {\bf \bar{x}}_i^{k}\|_L^2\right]^T$, we can arrive at the following relationship
from  (\ref{eq:Theorem_decreasing})  \as\, for all $k\ge0$:
\vspace{-0.15cm}
\be\label{eq-fin0}
\hspace{-0.2cm}\mathbb{E}\left[\bv^{k+1}|\mathcal{F}^k\right]
\le \left( \left[\begin{array}{cc}
1 &  \kappa_1\gamma^k \cr
0& 1-\kappa_2\gamma^k\cr\end{array}\right] +a^k {\bf 1}{\bf 1}^T \right)\bv^k+b^k{\bf 1}.
\vspace{-0.15cm}
\ee
By setting  $\pi=[1, \frac{\kappa_1}{\kappa_2}]^T$, we have
$
\pi^T \left[\begin{array}{cc}
1 &  \kappa_1\gamma^k  \cr
0& 1-\kappa_2\gamma^k\cr\end{array}\right]= \pi^T$.
Thus, relation~\eqref{eq-fin0} meets all   conditions of Lemma~\ref{le:vector_convergence}, implying that
  $\lim_{k\to\infty}\pi^T\bv^k$ exists \as, and that  $\{\sum_{i=1}^{m}\|\bar{x}_i^k-x_i^*\|_2^2\}$
and $\{\sum_{i=1}^m \|{\bf x}_i^{k} - {\bf \bar{x}}_i^{k}\|_L^2\}$ are bounded almost surely.

Consider the  second element of $\bv^k$ in \eqref{eq-fin0}, which should satisfy
$
 \mathbb{E}\left[\sum_{i=1}^m \|{\bf x}_i^{k+1} - {\bf \bar{x}}_i^{k+1}\|_L^2|\mathcal{F}^k\right]
 \le  (1-\kappa_2\gamma^k)\sum_{i=1}^m \|{\bf x}_i^{k} - {\bf \bar{x}}_i^{k}\|_L^2 + \b^k\quad\forall k\ge0,
$ \as, where
$\b^k\triangleq a^k\left( \sum_{i=1}^m\left(\|\bar{x}_i^k - x_i^\ast\|_2^2+ \|{\bf x}_i^{k} - {\bf \bar{x}}_i^{k}\|_L^2\right)\right)+b^k$.

Using the assumption that
$\sum_{k=0}^\infty a^k<\infty$ holds \as, and
the proven results that  $\{\sum_{i=1}^m\|\bar{x}_i^k-x_i^*\|_2^2\}$
and $\{\sum_{i=1}^m \|{\bf x}_i^{k} - {\bf \bar{x}}_i^{k}\|_L^2\}$ are bounded \as, one obtains  $\sum_{k=0}^\infty\b^k<\infty$ \as
 Thus, under the assumption of the proposition,   $\sum_{k=0}^\infty b^k<\infty$ \as\, and
$\sum_{k=0}^\infty \g^k=\infty$, $\sum_{i=1}^m \|{\bf x}_i^{k} - {\bf \bar{x}}_i^{k}\|_L^2$   satisfies the conditions of
Lemma~\ref{Lemma-polyak_2} with
  $q^k=\kappa_2\gamma^k$   and $p^k=\b^k$.
Therefore, we have the following relationship \as:
\be\label{eq-sumable}
 \sum_{k=0}^\infty \kappa_2\gamma^k\sum_{i=1}^m \|{\bf x}_i^{k} - {\bf \bar{x}}_i^{k}\|_L^2<\infty,\:
 \lim_{k\to\infty} \sum_{i=1}^m \|{\bf x}_i^{k} - {\bf \bar{x}}_i^{k}\|_L^2=0.
 \ee

We next proceed to prove $\sum_{i=1}^{m}\|\bar{x}_i^k-x_i^*\|^2\to0$ almost surely.
  One can verify that under $\sum_{k=0}^{\infty}a^k<\infty$ and $ \sum_{k=0}^{\infty}b^k<\infty $,  the inequality in (\ref{eq:Theorem_decreasing}) satisfies the relationship in Lemma \ref{le:vector_convergence} with $\bv^k=\left[\sum_{i=1}^{m}\|\bar{x}_i^k-x_i^*\|_2^2,\ \sum_{i=1}^m \|{\bf x}_i^{k} - {\bf \bar{x}}_i^{k}\|_L^2\right]^T$,   $V^k=\left[\begin{array}{cc}
1 &  \kappa_1\gamma^k \cr
0& 1-\kappa_2\gamma^k\cr
\end{array}\right]$, $H^k=\left[\begin{array}{cc}
c^k & 0 \cr
0& 0\cr
\end{array}\right]$, and  $\pi^T=[1, \frac{\kappa_1}{\kappa_2}]^T$. Therefore, from Lemma \ref{le:vector_convergence}, we arrive at the conclusion that $\pi^T\bv^k$ converges \as, i.e., $\sum_{i=1}^{m}\|\bar{x}_i^k-x_i^*\|_2^2+\frac{\kappa_1}{\kappa_2}\sum_{i=1}^m \|{\bf x}_i^{k} - {\bf \bar{x}}_i^{k}\|_L^2$ converges almost surely. Since  $\sum_{i=1}^m \|{\bf x}_i^{k} - {\bf \bar{x}}_i^{k}\|_L^2$ has been proven to converge  \as\, (see (\ref{eq-sumable})), we know that $\sum_{i=1}^{m}\|\bar{x}_i^k-x_i^*\|_2^2$ (equivalent to $\|\bar{x}^k-x^*\|_2^2$) converges almost surely.   Lemma \ref{le:vector_convergence} also implies $\sum_{ k=0 }^\infty \pi^TH^k\bu^k<\infty$   \as, i.e.,
$
\sum_{ k=0 }^\infty  \left[1, \frac{\kappa_1}{\kappa_2}\right]^T\hspace{-0.1cm}\left[\hspace{-0.1cm}\begin{array}{cc}
c^k & 0 \cr
0& 0\cr
\end{array}\hspace{-0.1cm}\right]\left[\hspace{-0.2cm}\begin{array}{c}
 \left(\phi(\bar{x}^k)- \phi(x^*)\right)^T(\bar{x}^k-x^{\ast}) \cr
 0\end{array}\hspace{-0.2cm}\right]\hspace{-0.1cm}<\hspace{-0.1cm}\infty,
$
or
\begin{equation}\label{eq:phi}
   \sum_{ k=0 }^\infty c^k\left(\phi(\bar{x}^k)- \phi(x^*)\right)^T(\bar{x}^k-x^{\ast})<\infty.
\end{equation}

Next, using (\ref{eq:phi}) and the proven \as\, convergence of $\|\bar{x}^k-x^{\ast}\|^2$, we   prove that $\bar{x}^k$ converges to $x^{\ast}$ almost surely.   The condition   $\sum_{k=0}^{\infty} c^k=\infty$, the property $\left(\phi(\bar{x}^k)-\phi(x^\ast)\right)^T(\bar{x}-x^\ast)>0$ (see Assumption \ref{as:Lipschitz}),  and (\ref{eq:phi}) imply that there exists a subsequence of $\{\bar{x}^k\}$, say $\{\bar{x}^{k_\ell}\}$, along which $\left(\phi(\bar{x}^k)- \phi(x^*)\right)^T(\bar{x}^k-x^{\ast})$ converges to zero almost surely.
The strongly monotone condition on  $\phi(\cdot)$   in  Assumption \ref{as:Lipschitz} implies that  $\{\bar{x}^{k_\ell}\}$ must converge to $x^{\ast}$ almost surely.
This and the fact that $ \|\bar{x}^k-x^{\ast}\|^2$ converges  \as\, imply that
 $\bar{x}^k$ converges to $x^{\ast}$ almost surely.
 Further note that $  \|{\bf x}_i^{k} - {\bf \bar{x}}_i^{k}\|_L^2$ converging to zero implies   $x_{(\ell)i}^k$ converging to $\bar{x}_i^k$ for all $\ell\in[m]$. Therefore, we have  $x_{(i)i}^k$ converging to $x_i^\ast$ \as\, for all $i\in[m]$.
\end{proof}

\section{Convergence Analysis for Algorithm 1}\label{sec:convergence_algorithm1}
In this section, based on Proposition~\ref{th-main_decreasing}, we
 establish the convergence of  Algorithm 1 to the unique NE under persistent DP  noise satisfying the following assumption:
\begin{Assumption 1}\label{assumption:dp-noise}
For every $i,\ell\in[m]$ and every $k$, conditional on the state $x_{(i)\ell}^k$,
the DP noise $\zeta_{(i)\ell}^k$ that player $i$ adds to its shared decision (or estimates of other players' decisions) satisfies $
\mathbb{E}\left[\zeta_{(i)\ell}^k\mid x_{(i)\ell}^k\right]=0$,  $\mathbb{E}\left[\|\zeta_{(i)\ell}^k\|^2\mid x_{(i)\ell}^k\right]=(\sigma_{i}^k)^2$, and 
$
\sum_{k=0}^\infty (\gamma^k)^2\, \max_{i\in[m]}(\sigma_{i}^k)^2 <\infty$,
  where $\{\gamma^k\}$ is from Algorithm 1.
Furthermore,
$\mathbb{E}\left[\|x_{(i)\ell}^0\|^2\right]<\infty$, $\forall i,\ell\in[m]$.
\end{Assumption 1}

\begin{Remark 1}
Since $\gamma^k$ decreases with time, Assumption \ref{assumption:dp-noise}  even allows the sequence $\{\sigma_i^k\}$ to increase  with time. For example, for $\gamma^k=\mathcal{O}(\frac{1}{k^{0.9}})$, if $\{\sigma_i^k\}$  increases with time with a rate no larger than $\mathcal{O}(k^{0.3})$, the summable condition   still holds. Allowing  $\{\sigma_i^k\}$ to  increase with time is key to enabling  strong $\epsilon$-DP, which will be detailed later in Theorem \ref{th:DP_Algorithm1}.
\end{Remark 1}

\begin{Theorem 1}\label{theorem:convergence_algorithm_1}
 Under Assumptions \ref{as:monotone}-\ref{assumption:dp-noise}, if there exists some $T\geq 0$ such that the sequences
  $\{\gamma^k\}$ and $\{\lambda^k\}$ satisfy
 \[
\sum_{k=T}^\infty \gamma^k=\infty, \:\sum_{k=T}^\infty \lambda^k=\infty, \: \sum_{k=T}^\infty (\gamma^k)^2<\infty,\:\sum_{k=T}^\infty \frac{(\lambda^k)^2}{\gamma^k}<\infty,
\]
 Algorithm~1 converges \as~to the unique NE of problem~(1).
\end{Theorem 1}

\begin{proof}
The basic idea  is to   prove that
$\sum_{i=1}^{m}\| \bar{x}_i^{k}-x_i^*\|^2$ and $\sum_{i=1}^{m}\|{\bf x}_i^{k}-{\bf \bar{x}}_i^{k}\|_L^2$ satisfy the conditions in Proposition \ref{th-main_decreasing}.

Part I: The evolution of  $\sum_{i=1}^{m}\|{\bf x}_i^{k}-{\bf \bar{x}}_i^{k}\|_L^2$.

From Algorithm 1, one can obtain the dynamics of ${\bf x}_i^k$:
\begin{equation}\label{eq:x_i_dynamics}
{\bf x}_i^{k+1}=(I+\gamma^kL){\bf x}_i^k+\gamma^kL_o\bfs{\zeta}_i^k-\lambda^ke_i F_i^T(x_{(i)i}^k,x_{(i)-i}^k),
\end{equation}
where $e_i\in\mathbb{R}^{m}$ is a unitary vector with the $i$th element equal to $1$ and all the other elements equal to zero, $L_o\in\mathbb{R}^{m\times m}$ is the matrix obtained by replacing all diagonal entries of matrix $L$ with zero, and
$
\bfs{\zeta}_i^k=  [ \zeta_{(1)i}^k, \cdots   \zeta_{(m)i}^k  ]^T\in\mathbb{R}^{m\times d_i}
$.

One can obtain that $
{\bf \bar{x}}_i^k =\frac{{\bf 1}u^T {\bf x}_i^k}{m}
$ always holds, which, in combination with (\ref{eq:x_i_dynamics}), yields
\begin{equation}\label{eq:bar_x_i_dynamics}
\begin{aligned}
{\bf \bar{x}}_i^{k+1}
&\hspace{-0.1cm}={\bf \bar{x}}_i^{k}+\textstyle\gamma^k\frac{{\bf 1}u^TL_o}{m}\bfs{\zeta}_i^k-\frac{\lambda^k{\bf 1}u_i F_i^T(x_{(i)i}^k,x_{(i)-i}^k)}{m},
\end{aligned}
\end{equation}
where  $u_i$ is the $i$th entry of $u$.
Note that in the last equality we used the property $u^T(I+\gamma^kL)=u^T$ from Lemma \ref{lemma:coupling}.

Combining (\ref{eq:x_i_dynamics}) with (\ref{eq:bar_x_i_dynamics}) yields
\begin{equation}\label{eq:x_i_dynamics3}
\begin{aligned}
 {\bf x}_i^{k+1}-{\bf \bar{x}}_i^{k+1}&=W^k{\bf x}_i^k+\gamma^k\Pi_{L_o}\bfs{\zeta}_i^k
 -\lambda^k\Pi_{e_i} F_i^T(x_{(i)i}^k,x_{(i)-i}^k),
\end{aligned}
\end{equation}
where we have defined
$
W^k\triangleq I+\gamma^kL-\frac{{\bf 1}u^T}{m}$, $\Pi_{L_o}\triangleq { L_o}-\frac{{\bf 1}u^TL_o}{m}$, and $\Pi_{e_i}\triangleq e_i-\frac{u_i{\bf 1}}{m}$.

The second statement of Lemma \ref{lemma:coupling} implies $W^k{\bf 1}=0$ and further $W^k{\bf \bar{x}}_i^{k}=0$. Hence, we can subtract $W^k{\bf \bar{x}}_i^{k}=0$ from the right hand side of (\ref{eq:x_i_dynamics3}) to obtain
\begin{equation}\label{eq:x_i_dynamics4}
\begin{aligned}
 {\bf x}_i^{k+1}\hspace{-0.15cm}-\hspace{-0.05cm} {\bf \bar{x}}_i^{k+1} \hspace{-0.09cm}=\hspace{-0.05cm}W^k\hspace{-0.05cm}({\bf x}_i^k\hspace{-0.05cm}-\hspace{-0.05cm}{\bf \bar{x}}_i^{k})\hspace{-0.05cm}+\hspace{-0.05cm}\gamma^k\Pi_{L_o}\hspace{-0.05cm}\bfs{\zeta}_i^k
\hspace{-0.15cm}-\hspace{-0.05cm}\lambda^k\hspace{-0.03cm}\Pi_{e_i} \hspace{-0.05cm}F_i^T\hspace{-0.08cm}(x_{\hspace{-0.05cm}(i)i}^k,\hspace{-0.05cm}x_{\hspace{-0.05cm}(i)-i}^k\hspace{-0.03cm}).
\end{aligned}
\end{equation}
Taking the $\|\cdot\|_L$ norm on both sides leads to
\begin{equation}\label{eq:x_i_-barx_i}
\begin{aligned}
& \|{\bf x}_i^{k+1}\hspace{-0.15 cm}-\hspace{-0.05cm} {\bf \bar{x}}_i^{k+1}\|_L^2\leq\|W^k({\bf x}_i^k \hspace{-0.05cm}-\hspace{-0.05cm} {\bf \bar{x}}_i^{k}) \hspace{-0.05cm}-\hspace{-0.05cm}\lambda^k\Pi_{e_i} F_i^T(x_{(i)i}^k,x_{(i)-i}^k)\|_L^2\\
 &\:+(\gamma^k)^2\|\Pi_{L_o}\|_L^2\|\bfs{\zeta}_i^k\|_L^2\\
 &\: +2\left\langle W^k({\bf x}_i^k- {\bf \bar{x}}_i^{k})-\lambda^k\Pi_{e_i} F_i^T(x_{(i)i}^k,x_{(i)-i}^k),\gamma^k\Pi_{L_o}\bfs{\zeta}_i^k\right\rangle_L.
\end{aligned}
\end{equation}

Taking the conditional expectation on both sides, with
respect to $\mathcal{F}_k=\{{\bf x}_i^\ell; 0\leq \ell\leq k,\,i\in[m]\}$, leads to
\begin{equation}\label{eq:x_i_-barx_i_2}
\begin{aligned}
&\mathbb{E}\left[ \|{\bf x}_i^{k+1}- {\bf \bar{x}}_i^{k+1}\|_L^2|\mathcal{F}^k\right] \hspace{-0.05cm}\leq \hspace{-0.05cm} (\gamma^k)^2\|\Pi_{L_o}\|_L^2\delta_{L,2}^2  m(\sigma_i^k)^2 \\
&+\|W^k({\bf x}_i^k- {\bf \bar{x}}_i^{k})-\lambda^k\Pi_{e_i} F_i^T(x_{(i)i}^k,x_{(i)-i}^k)\|_L^2,
 \end{aligned}
\end{equation}
where we have used  Assumption \ref{assumption:dp-noise}  and the property   $\|\bfs{\zeta}_i^k\|_L^2\leq \delta_{L,2}^2\|\bfs{\zeta}_i^k\|_2^2$ (see Lemma \ref{lemma:normR_norm2}).

Further  using   Lemma \ref{lemma:normR_norm2}, we can simplify (\ref{eq:x_i_-barx_i_2}) as
\begin{equation}\label{eq:x_i_-barx_i_3}
\begin{aligned}
&\mathbb{E}\left[ \|{\bf x}_i^{k+1}- {\bf \bar{x}}_i^{k+1}\|_L^2|\mathcal{F}^k\right] \hspace{-0.10cm}\leq \hspace{-0.10cm}(\gamma^k)^2\|\Pi_{L_o}\|_L^2\delta_{L,2}^2  m(\sigma_i^k)^2\\
 &+\hspace{-0.10cm}\left(\|W^k\|_L \|{\bf x}_i^k- {\bf \bar{x}}_i^{k}\|_L \hspace{-0.10cm}+\hspace{-0.10cm}   \lambda^k \|\Pi_{e_i}\|_L  \|F_i^T(x_{(i)i}^k,x_{(i)-i}^k) \|_L\right)^2.
 \end{aligned}
\end{equation}

According to Lemma \ref{lemma:coupling}, we have $\|W^k\|_L\leq 1-\alpha\gamma^k$ for some $0<\alpha<1$ when $\gamma^k$ is sufficiently  small. Given that $\{\gamma^k\}$ is square summable, we have $\|W^k\|_L\leq 1-\alpha\gamma^k$ for some $0<\alpha<1$ when $k$ is larger than some $T$. Therefore, (\ref{eq:x_i_-barx_i_3}) means that there always exists a $T\geq 0$ such that we   have
\begin{equation}\label{eq:x_i_-barx_i_4}
\begin{aligned}
&\mathbb{E}\left[ \|{\bf x}_i^{k+1}- {\bf \bar{x}}_i^{k+1}\|_L^2|\mathcal{F}^k\right]\leq(\gamma^k)^2\|\Pi_{L_o}\|_L^2\delta_{L,2}^2  m(\sigma_i^k)^2\\
 &+\left((1-\alpha\gamma^k) \|{\bf x}_i^k- {\bf \bar{x}}_i^{k}\|_L \hspace{-0.1cm}+ \hspace{-0.1cm}  \lambda^k \|\Pi_{e_i}\|_L  \|F_i^T(x_{(i)i}^k,x_{(i)-i}^k) \|_L\right)^2
 \end{aligned}
\end{equation}
for $k\geq T$.

Applying to the second term on the right hand side of (\ref{eq:x_i_-barx_i_4}) the inequality $(a+b)^2\le (1+\epsilon) a^2 + (1+\epsilon^{-1})b^2$, valid for any scalars $a,b,$ and $\epsilon>0$, we can obtain
the following relationship by setting $\epsilon$ as $\frac{\gamma^k\alpha}{1-\gamma^k\alpha}$ (which further results in $1+\epsilon=\frac{1}{1-\gamma^k\alpha}$ and $1+\epsilon^{-1}=\frac{1}{\gamma^k\alpha}$):
\begin{equation}\label{eq:x_i_-barx_i_6}
\begin{aligned}
&\mathbb{E}\left[ \|{\bf x}_i^{k+1}- {\bf \bar{x}}_i^{k+1}\|_L^2|\mathcal{F}^k\right]\leq(\gamma^k)^2\|\Pi_{L_o}\|_L^2\delta_{L,2}^2  m(\sigma_i^k)^2\\
 &+\hspace{-0.05cm}(1\hspace{-0.05cm}-\hspace{-0.05cm}\alpha\gamma^k) \|{\bf x}_i^k\hspace{-0.05cm}-\hspace{-0.05cm} {\bf \bar{x}}_i^{k}\|^2_L   + \frac{(\lambda^k)^2 \|\Pi_{e_i}\|^2_L }{\gamma^k\alpha} \|F^T_i(x_{(i)i}^k,x_{(i)-i}^k) \|^2_L.
 \end{aligned}
\end{equation}

Next, we use Assumption \ref{as:Lipschitz} to bound $\|F^T_i(x_{(i)i}^k,x_{(i)-i}^k) \|^2_L$.

At the NE point $x^{\ast}$,  we always have
$F_i(x_i^\ast,x^\ast_{-i})=0$ for all $i\in[m]$, which implies
\begin{equation}\label{eq:gradient_bound}
\begin{aligned}
& \|F_i(x_{(i)i}^k,x_{(i)-i}^k) \|^2_L\\
&\leq\delta^2_{L,2}\|F_i(x_{(i)i}^k,x_{(i)-i}^k)- F_i(x_{(i)i}^k,x_{-i}^\ast)+F_i(x_{(i)i}^k,x_{-i}^\ast) \\
&\qquad -F_i(x_i^\ast,x^\ast_{-i})\|^2_2\\
&\leq 2{ K_2^2}\delta^2_{L,2}\|x_{(i)-i}^k- x_{-i}^\ast\|_2^2   +2{ K_1^2}\delta^2_{L,2}\| x_{(i)i}^k- x_i^\ast \|^2_2,
 \end{aligned}
\end{equation}
where in the last inequality we used Assumption \ref{as:Lipschitz}.

Using inequalities $ \|x_{(i)-i}^k- x_{-i}^\ast\|_2^2 
 \leq 2\|x_{(i)-i}^k-\bar{x}_{-i}^k\|_2^2+2\|\bar{x}_{-i}^k-x_{-i}^\ast\|_2^2
  $ and $\|x_{(i)i}^k- x_i^\ast\|_2^2
  \leq 2\|x_{(i)i}^k-\bar{x}_{i}^k\|_2^2+2\|\bar{x}_{i}^k-x_{i}^\ast\|_2^2$, where $\bar{x}_{-i}^k\triangleq[(\bar{x}_1^k)^T,\cdots,(\bar{x}_{i-1}^k)^T,(\bar{x}_{i+1}^k)^T,\cdots,(\bar{x}_m^k)^T]^T$,
we can obtain
\begin{equation}\label{eq:gradient_bound2}
\begin{aligned}
& \|F_i(x_{(i)i}^k,x_{(i)-i}^k) \|^2_L \leq 4{ K_1^2}\delta^2_{L,2}(\|x_{(i)i}^k-\bar{x}_{i}^k\|_2^2+ \|\bar{x}_{i}^k-x_{i}^\ast\|_2^2) \\
&           + 4{ K_2^2} \delta^2_{L,2}(\|x_{(i)-i}^k-\bar{x}_{-i}^k\|_2^2+ \|\bar{x}_{-i}^k-x_{-i}^\ast\|_2^2).
 \end{aligned}
\end{equation}

Plugging (\ref{eq:gradient_bound2}) into  (\ref{eq:x_i_-barx_i_6}) leads to
\begin{equation}\label{eq:x_i_-barx_i_62}
\begin{aligned}
&\mathbb{E}\left[ \|{\bf x}_i^{k+1}- {\bf \bar{x}}_i^{k+1}\|_L^2|\mathcal{F}^k\right]\\
 &\leq(1-\alpha\gamma^k) \|{\bf x}_i^k- {\bf \bar{x}}_i^{k}\|^2_L +(\gamma^k)^2\|\Pi_{L_o}\|_L^2\delta_{L,2}^2  m(\sigma_i^k)^2\\
&
+ \textstyle\frac{4(\lambda^k)^2 \|\Pi_{e_i}\|^2_L { K_2^2} \delta^2_{L,2}}{\gamma^k\alpha}(\|x_{(i)-i}^k-\bar{x}_{-i}^k\|_2^2 +\|\bar{x}_{-i}^k-x_{-i}^\ast\|_2^2)\\
&
+ \textstyle\frac{4(\lambda^k)^2 \|\Pi_{e_i}\|^2_L { K_1^2}\delta^2_{L,2}}{\gamma^k\alpha}(\|x_{(i)i}^k-\bar{x}_{i}^k\|_2^2+\|\bar{x}_{i}^k-x_{i}^\ast\|_2^2).
 \end{aligned}
\end{equation}
Summing (\ref{eq:x_i_-barx_i_62}) from $i=1$ to $i=m$
 and noting
$
\sum_{i=1}^{m}\|{\bf x}_i^k- {\bf \bar{x}}_i^k\|_2^2=\sum_{i=1}^{m} (\|x_{(i)i}^k-\bar{x}_{i}^k\|_2^2+\|x_{(i)-i}^k-\bar{x}_{-i}^k\|_2^2 )
$
and
$
\sum_{i=1}^{m}\|\bar{x}_{-i}^k-x_{-i}^\ast\|_2^2=(m-1)\sum_{i=1}^{m}\|\bar{x}_{i}^k-x_{i}^\ast\|_2^2
$,
we  obtain
 \begin{equation}\label{eq:x_i_-barx_i_8}
\begin{aligned}
&\textstyle\mathbb{E}\left[ \sum_{i=1}^{m}\|{\bf x}_i^{k+1}\hspace{-0.1cm}-\hspace{-0.1cm} {\bf \bar{x}}_i^{k+1}\|_L^2|\mathcal{F}^k\right] \hspace{-0.05cm}\leq \hspace{-0.05cm}(1\hspace{-0.05cm}-\hspace{-0.05cm}\alpha\gamma^k) \sum_{i=1}^{m}\|{\bf x}_i^k- {\bf \bar{x}}_i^{k}\|^2_L\\
 & +\textstyle(\gamma^k)^2\|\Pi_{L_o}\|_L^2\delta_{L,2}^2  m\sum_{i=1}^{m}(\sigma_i^k)^2\\
 &+  \textstyle\frac{4m(\lambda^k)^2 \|\Pi_{e_i}\|^2_L  \tilde{K}^2\delta^2_{L,2}}{\gamma^k\alpha} \sum_{i=1}^{m}\|\bar{x}_{i}^k-x_{i}^\ast\|_2^2\\
&
+ \textstyle\frac{4(\lambda^k)^2 \|\Pi_{e_i}\|^2_L  \tilde{K}^2 \delta^2_{L,2}\delta^2_{2,L}}{\gamma^k\alpha}\sum_{i=1}^{m}\|{\bf x}_i^k- {\bf \bar{x}}_i^{k}\|^2_L,
 \end{aligned}
\end{equation}
where we have defined $\tilde{K}\triangleq\max\{{ K_1},{ K_2}\}$.

Part II: The evolution of $\sum_{i=1}^{m}\| \bar{x}_i^{k}-x_i^*\|^2$.

Subtracting $x_i^\ast$ from both sides of (\ref{eq:bar_x_i_dynamics}) yields
 \begin{equation}
\begin{aligned}
&\|\bar{x}_i^{k+1}-x_i^\ast\|_2^2
 \leq \|\bar{x}_i^{k}-x_i^\ast\|^2+2(\gamma^k)^2\textstyle\frac{\|u^TL_o\|_2^2}{m^2}\|\bfs{\zeta}_i^k\|_2^2\\
&\quad +\textstyle\frac{2(\lambda^k)^2u_i^2}{m^2}\|F_i(x_{(i)i}^k,x_{(i)-i}^k)\|_2^2\\
&\quad\textstyle +2\left\langle\bar{x}_i^{k}-x_i^\ast,\gamma^k\frac{(u^TL_o\bfs{\zeta}_i^k)^T}{m}-\frac{1}{m}\lambda^ku_i F_i(x_{(i)i}^k,x_{(i)-i}^k) \right\rangle.
\end{aligned}
\end{equation}
Taking the conditional expectation on both sides, with
respect to $\mathcal{F}_k=\{{\bf x}_i^\ell; 0\leq \ell\leq k,\,i\in[m]\}$, leads to
\begin{equation}\label{eq:bar_x-x_ast}
\begin{aligned}
&\mathbb{E}\left[\|\bar{x}_i^{k+1}-x_i^\ast\|_2^2|\mathcal{F}^k\right]\leq \|\bar{x}_i^{k}-x_i^\ast\|_2^2+\textstyle\frac{2(\gamma^k)^2\|u^TL_o\|_2^2(\sigma_i^k)^2}{m}\\
& +\textstyle\frac{2(\lambda^k)^2u_i^2\|F_i(x_{(i)i}^k,x_{(i)-i}^k)\|_2^2}{m^2}  -\textstyle\frac{2 u_i \lambda^k \langle\bar{x}_i^{k}-x_i^\ast,  F_i(x_{(i)i}^k,x_{(i)-i}^k)  \rangle}{m}.
\end{aligned}
\end{equation}

Next we bound the last two terms in (\ref{eq:bar_x-x_ast}).
For the second last term, we can bound it similarly to (\ref{eq:gradient_bound2}):
\begin{equation}\label{eq:gradient_bound3}
\begin{aligned}
& \|F_i(x_{(i)i}^k,x_{(i)-i}^k) \|^2_2\leq 4{ K_2^2}  \|x_{(i)-i}^k-\bar{x}_{-i}^k\|_2^2\\
&+4{ K_2^2} \|\bar{x}_{-i}^k-x_{-i}^\ast\|_2^2      +4{ K_1^2} \|x_{(i)i}^k-\bar{x}_{i}^k\|_2^2+4{ K_1^2} \|\bar{x}_{i}^k-x_{i}^\ast\|_2^2.
 \end{aligned}
\end{equation}

For the inner-product term in (\ref{eq:bar_x-x_ast}), we bound it using  $F_i(x_i^\ast,x^\ast_{-i})=0$ and split it as follows:
\begin{equation}\label{eq:inner-product}
\begin{aligned}
& 2\lambda^k\left\langle\bar{x}_i^{k}-x_i^\ast,  F_i(x_{(i)i}^k,x_{(i)-i}^k)-F_i(\bar{x}_i^k,\bar{x}_{-i}^k) \right\rangle \\
&  +2\lambda^k\left\langle\bar{x}_i^{k}-x_i^\ast,  F_i(\bar{x}_i^k,\bar{x}_{-i}^k)-F_i(x_i^\ast,x^\ast_{-i}) \right\rangle.
\end{aligned}
\end{equation}
For the  first inner-product term on the right hand side of (\ref{eq:inner-product}), using the Cauchy-Schwarz inequality  yields
\begin{equation}\label{eq:inner_product_sub2}
\begin{aligned}
 &2\lambda^k\left\langle\bar{x}_i^{k}-x_i^\ast,  F_i(x_{(i)i}^k,x_{(i)-i}^k)-F_i(\bar{x}_i^k,\bar{x}_{-i}^k) \right\rangle\\
 &\geq -\textstyle\frac{(\lambda^k)^2{ \|\bar{x}_i^{k}-x_i^\ast\|^2_2}}{\gamma^k}-\gamma^k\|F_i(x_{(i)i}^k,x_{(i)-i}^k)-F_i(\bar{x}_i^k,\bar{x}_{-i}^k)\|_2^2.
 \end{aligned}
\end{equation}
The Lipschitz assumption in Assumption \ref{as:Lipschitz} implies
\begin{equation}\label{eq:inner_product2}
\begin{aligned}
&\|F_i(x_{(i)i}^k,x_{(i)-i}^k)-F_i(\bar{x}_i^k,\bar{x}_{-i}^k)\|^2_2\\
&\leq 2\|F_i(x_{(i)i}^k,x_{(i)-i}^k)-F_i(\bar{x}_{i}^k,x_{(i)-i}^k)\|_2^2\\
&\quad +2\|F_i(\bar{x}_{i}^k,x_{(i)-i}^k)- F_i(\bar{x}_i^k,\bar{x}_{-i}^k)\|^2_2\\
&\leq 2  {K}^2_1\|x_{(i)i}^k-\bar{x}_{i}^k\|_2^2 +2  {K}^2_2\|x_{(i)-i}^k -  \bar{x}_{-i}^k\|^2_2.
\end{aligned}
\end{equation}

Combining (\ref{eq:inner-product}), (\ref{eq:inner_product_sub2}), and (\ref{eq:inner_product2}) leads to
\begin{equation}\label{eq:inner-product_final}
\begin{aligned}
&2\lambda^k\left\langle\bar{x}_i^{k}-x_i^\ast,  F_i(x_{(i)i}^k,x_{(i)-i}^k) \right\rangle\\
&\geq - \textstyle\frac{(\lambda^k)^2{ \|\bar{x}_i^{k}-x_i^\ast\|^2_2}}{\gamma^k}-2 {K}^2_1\gamma^k\|x_{(i)i}^k-\bar{x}_{i}^k\|_2^2\\
&\quad -2  {K}^2_2\gamma^k\|x_{(i)-i}^k -  \bar{x}_{-i}^k\|^2_2\\
&\quad + 2\lambda^k\left\langle\bar{x}_i^{k}-x_i^\ast,  F_i(\bar{x}_i^k,\bar{x}_{-i}^k)-F_i(x_i^\ast,x^\ast_{-i}) \right\rangle.
\end{aligned}
\end{equation}

Further substituting (\ref{eq:gradient_bound3}) and (\ref{eq:inner-product_final}) into (\ref{eq:bar_x-x_ast}) yields
\begin{equation}\label{eq:bar_x-x_ast2}
\begin{aligned}
&\mathbb{E}\left[\|\bar{x}_i^{k+1}-x_i^\ast\|_2^2|\mathcal{F}^k\right] \leq \|\bar{x}_i^{k}-x_i^\ast\|_2^2+2(\gamma^k)^2\textstyle\frac{\|u^TL_o\|_2^2}{m}(\sigma_i^k)^2\\
& +\textstyle\frac{8(\lambda^k)^2u_i^2 { K_2^2}}{m^2}  (\|x_{(i)-i}^k-\bar{x}_{-i}^k\|_2^2 + \|\bar{x}_{-i}^k-x_{-i}^\ast\|_2^2)\\
&    +\textstyle\frac{8(\lambda^k)^2u_i^2{ K_1^2}}{m^2} (\|x_{(i)i}^k-\bar{x}_{i}^k\|_2^2+ \|\bar{x}_{i}^k-x_{i}^\ast\|_2^2)\\
& + \textstyle\frac{u_i(\lambda^k)^2{ \|\bar{x}_i^{k}-x_i^\ast\|^2_2}}{m\gamma^k}+\frac{2u_i {K}^2_1\gamma^k}{m}\|x_{(i)i}^k-\bar{x}_{i}^k\|_2^2\\
& +\textstyle\frac{2u_i  {K}^2_2\gamma^k}{m}\|x_{(i)-i}^k -  \bar{x}_{-i}^k\|^2_2\\
& -\textstyle\frac{2u_i\lambda^k}{m}\left\langle\bar{x}_i^{k}-x_i^\ast,  F_i(\bar{x}_i^k,\bar{x}_{-i}^k)-F_i(x_i^\ast,x^\ast_{-i}) \right\rangle.
\end{aligned}
\end{equation}

Summing (\ref{eq:bar_x-x_ast2}) from $i=1$ to $i=m$, and using the relationship
$
\sum_{i=1}^{m}\|\bar{x}_{-i}^k-x_{-i}^\ast\|_2^2=(m-1)\sum_{i=1}^{m}\|\bar{x}_{i}^k-x_{i}^\ast\|_2^2
$ and $
\sum_{i=1}^{m}\|{\bf x}_i^k- {\bf \bar{x}}_i^k\|_2^2=\sum_{i=1}^{m}\left(\|x_{(i)i}^k-\bar{x}_{i}^k\|_2^2+\|x_{(i)-i}^k-\bar{x}_{-i}^k\|_2^2\right)$
 lead to
\begin{equation}\label{eq:bar_x-x_ast_final}
\begin{aligned}
&\textstyle\mathbb{E}\left[\sum_{i=1}^{m}\|\bar{x}_i^{k+1}-x_i^\ast\|_2^2|\mathcal{F}^k\right]\\
&\leq\textstyle \sum_{i=1}^{m}\|\bar{x}_i^{k}-x_i^\ast\|_2^2+2(\gamma^k)^2\frac{\|u^TL_o\|_2^2}{m}\sum_{i=1}^{m}(\sigma_i^k)^2\\
&+\textstyle\frac{8(\lambda^k)^2u_i^2  \tilde{K}^2}{m^2} \sum_{i=1}^{m} \|{\bf x}_{i}^k-\bar{\bf x}_{i}^k\|_2^2 +\frac{ 8(\lambda^k)^2u_i^2 \tilde{K}^2}{m} \sum_{i=1}^{m}\|\bar{x}_{i}^k-x_{i}^\ast\|_2^2\\
& +\textstyle \frac{u_i(\lambda^k)^2\sum_{i=1}^{m}{ \|\bar{x}_i^{k}-x_i^\ast\|^2_2}}{m\gamma^k} + \frac{2u_i \tilde{K}^2 \gamma^k}{m} \sum_{i=1}^{m}\|{\bf x}_{i}^k-\bar{\bf x}_{i}^k\|^2_2\\
&-\textstyle\frac{2u_i\lambda^k}{m} \left(\phi(\bar{x}^k)-\phi(x^\ast)\right)^T(\bar{x}^k-x^\ast),
\end{aligned}
\end{equation}
where $\bar{x}^k=\left[(\bar{x}_1^k)^T,\,\cdots,(\bar{x}_m^k)^T\right]^T$ and $ \tilde{K}\triangleq\max\{{ K_1},{ K_2}\}$.

Part III: Combination of Step I and Step II.

By combining (\ref{eq:x_i_-barx_i_8}) and (\ref{eq:bar_x-x_ast_final}), and using Assumption \ref{assumption:dp-noise}, we have    $\sum_{i=1}^{m}\|{\bf x}_{i}^k-\bar{\bf x}_{i}^k\|_L^2$ and $\sum_{i=1}^{m}\|\bar{x}_i^{k}-x_i^\ast\|_2^2$ satisfying the conditions in Proposition \ref{th-main_decreasing} with $\kappa_1=\frac{2u_i \tilde{K}^2 \delta^2_{L,2}}{m}$, $\kappa_2=\alpha$, $a^k=\max\{a_1^k,\, a_2^k,  a_3^k, a_4^k, a_5^k\}$,
$a_1^k\triangleq \frac{4(\lambda^k)^2 \|\Pi_{e_i}\|^2_L  \tilde{K}^2 \delta^2_{L,2}\delta^2_{2,L}}{\gamma^k\alpha}$, $a_2^k\triangleq \frac{4m(\lambda^k)^2 \|\Pi_{e_i}\|^2_L  \tilde{K}^2\delta^2_{L,2}}{\gamma^k\alpha}$,
$a_3^k\triangleq \frac{8(\lambda^k)^2u_i^2  \tilde{K}^2\delta_{L,2}^2}{m^2}$,
$a_4^k\triangleq \frac{8 (\lambda^k)^2u_i^2 \tilde{K}^2}{m}$, $a_5^k\triangleq\frac{u_i(\lambda^k)^2}{m\gamma^k}$,
 $b^k=\max\{b_1^k,b_2^k\}$, $b_1^k\triangleq (\gamma^k)^2\|\Pi_{L_o}\|_L^2\delta_{L,2}^2  m\sum_{i=1}^{m}(\sigma_i^k)^2$, $b_2^k\triangleq 2(\gamma^k)^2\frac{\|u^TL_o\|_2^2}{m}\sum_{i=1}^{m}(\sigma_i^k)^2$, and $c^k=\frac{2u_i\lambda^k}{m}$.
\end{proof}

{
\begin{Remark 1}
  The requirement on  $\gamma^k$ and  $\lambda^k$ in the statement  of Theorem \ref{theorem:convergence_algorithm_1} can be satisfied, for example, by setting $\gamma^k=\mathcal{O}(\frac{1}{k^a})$ and $\lambda^k=\mathcal{O}(\frac{1}{k^b})$ with $a,b\in\mathbb{R}$ satisfying $0.5<a<b\leq 1$ and $2b-a>1$. For example, setting $\gamma^k=\frac{c_1}{1+c_2k^{\iota}}$ and $\lambda^k=\frac{c_3}{1+c_4k}$ will satisfy the conditions for any   exponent $0.5<\iota<1$, and positive coefficients   $c_1$, $c_2$, $c_3$, and $c_4$.
\end{Remark 1}
}
\begin{Remark 1}
  Leveraging Proposition \ref{th-main_decreasing}, we can also analyze the convergence speed, the details of which are given in the extended version in \cite{wang2022ensuring}.
\end{Remark 1}

\begin{Remark 1}\label{Re-rate}
Since  $x_{(i)i}^k$ converges to the NE following the dynamics (\ref{eq:Theorem_decreasing}) in Proposition \ref{th-main_decreasing}, we  leverage  (\ref{eq:Theorem_decreasing}) in Proposition \ref{th-main_decreasing} to characterize the convergence speed.
 The first relationship in (\ref{eq-sumable}) (i.e., $\sum_{k=0}^\infty \kappa_2\gamma^k\sum_{i=1}^m \|{\bf x}_i^{k} - {\bf \bar{x}}_i^{k}\|_L^2<\infty$)  implies that $\|{\bf x}_i^{k} - {\bf \bar{x}}_i^{k}\|_L^2$ decreases to zero no slower than $\mathcal{O}(\frac{1}{k\gamma^k})$, implying that the convergence speed of ${\bf x}_i^k$  to ${\bf \bar{x}}_i^k$ is no slower than $\mathcal{O}(\frac{1}{(k\gamma^k)^{0.5}})$. Therefore, the convergence speed of the decision variable $x_{(i)i}^k$ to the average $\bar{x}_{i}^k$ is no slower than $\mathcal{O}(\frac{1}{(k\gamma^k)^{0.5}})$. Eqn. \eqref{eq:Theorem_decreasing}  implies the following inequality:
\begin{equation}
\begin{aligned}
&\mathbb{E}\left[\sum_{i=1}^{m}\|\bar{x}_i^{k+1}-x_i^*\|_2^2|\mathcal{F}^k\right]
\le  (1+a^k)\sum_{i=1}^{m}\|\bar x_i^k -x_i^*\|_2^2 \cr
&+\left( \kappa_1\gamma^k +a^k\right) \sum_{i=1}^m \|{\bf x}_i^k - {\bf \bar{x}}_i^k\|_L^2 + b^k.
\end{aligned}
\end{equation}
Given  $\left( \kappa_1\gamma^k +a^k\right)\sum_{i=1}^m \|{\bf x}_i^{k} - {\bf \bar{x}}_i^{k}\|_L^2<\infty$ (note that  $\sum_{k=0}^{\infty}a^k<\infty$ and $\sum_{k=0}^{\infty}\gamma^k=\infty$ imply that $a^k$ decreases faster than $\{\gamma^k\}$), the following relationship holds \as:
 \begin{equation}
\begin{aligned}
&\mathbb{E}\left[\sum_{i=1}^{m}\|\bar{x}_i^{k+1}-x_i^*\|_2^2|\mathcal{F}^k\right]
\le  (1+a^k)\sum_{i=1}^{m}\|\bar x_i^k -x_i^*\|_2^2  +  \hat{b}^k,
\end{aligned}
\end{equation}
where   $\hat{b}^k\triangleq\left( \kappa_1\gamma^k +a^k\right) \sum_{i=1}^m \|{\bf x}_i^k - {\bf \bar{x}}_i^k\|_L^2 + b^k$ satisfies $\sum_{k=0}^{\infty}\hat{b}^k<\infty$.
Since  non-negative summable sequences decrease to zero no slower than $\mathcal{O}(\frac{1}{k})$, we know that  $\mathbb{E}\left[\sum_{i=1}^{m}\|\bar{x}_i^{k+1}-x_i^*\|^2\right]$ decreases to zero no slower than  $\mathcal{O}(\frac{1}{k})$.
Therefore, the convergence of every decision variable $x_{(i)i}^k$ to the NE $x_i^\ast$, which is the combination of the convergence of $x_{(i)i}^k$ to $\bar{x}_i^k$ and the convergence of $\bar{x}_i^k$ to $x_i^\ast$, is no slower than $\mathcal{O}(\frac{1}{(k\gamma^k)^{0.5}})$. (For example, under $\gamma^k=\mathcal{O}(\frac{1}{k^{0.6}})$, $\mathcal{O}(\frac{1}{(k\gamma^k)^{0.5}})$ is $\mathcal{O}(\frac{1}{k^{0.2}})$.)
\end{Remark 1}

\section{Privacy Analysis of Algorithm 1}\label{se:privacy_Algorithm1}
\begin{Definition 1}\label{de:sensitivity}
  For  any  given initial state $\vartheta^0$ and adjacent networked games  $\mathcal{P}$ and $\mathcal{P'}$,  the sensitivity of an NE seeking algorithm at iteration $k$ is
  \begin{equation}
 \hspace{-0.3cm} \Delta^k\triangleq \sup\limits_{\mathcal{O}\in\mathbb{O}}\left\{\sup\limits_{\vartheta\in{ \mathcal{R}_{\mathcal{P},\vartheta^0}^{-1}}(\mathcal{O}),\:\vartheta'\in{ \mathcal{R}_{\mathcal{P'},\vartheta^0}^{-1}}(\mathcal{O})}\hspace{-0.35cm}\|\vartheta^{k+1}-\vartheta'^{k+1}\|_1\right\}.
  \end{equation}
\end{Definition 1}

Based on this definition, we obtain the following result:
\begin{Lemma 1}\label{Le:Laplacian}
At each iteration $k$, if each player in Algorithm 1  adds a vector noise  $\zeta_{(i)\ell}^k\in\mathbb{R}^{d_i}$ (consisting of $d_i$ independent Laplace noises with  parameter $\nu^k$) to each of its shared message $x_{(i)\ell}^k$ such that $\sum_{k=1}^{T_0}\frac{\Delta^k}{\nu^k}\leq \bar\epsilon$, then  Algorithm 1 is $\epsilon$-differentially private with the cumulative privacy budget from iterations  $k=0$ to $k=T_0$ less than $\bar\epsilon$.
\end{Lemma 1}
\begin{proof}
The result can be obtained following the derivation of Lemma 2 in  \cite{Huang15} (see also Theorem 3 in \cite{ye2021differentially}), and hence we do not include  the proof here.
\end{proof}

As is the case in \cite{Huang15}, since the change in  $f_i$ can be arbitrarily large in  Definition \ref{de:adjacency}, we have to introduce the following assumption to ensure a bounded sensitivity:
\begin{Assumption 1}\label{as:bounded_gradients}
  $\|F_i(x_{(i)i},x_{(i)-i})\|_1\leq \bar{C}$ holds for all $x_{(i)i}\in\mathbb{R}^{d_i}$, $x_{(i)-i}\in\mathbb{R}^{D-d_i}$, and $1\leq i \leq m$, where $\bar{C}$   is a constant.
\end{Assumption 1}
\begin{Theorem 1}\label{th:DP_Algorithm1}
Under Assumptions \ref{as:monotone}, \ref{as:Lipschitz}, \ref{Assumption:coupling}, \ref{as:bounded_gradients}, if   $\{\lambda^k\}$ and $\{\gamma^k\}$ satisfy the conditions in Theorem \ref{theorem:convergence_algorithm_1}, and all elements of $\zeta_{(i)1}^k,\,\cdots,\,\zeta_{(i)m}^k$ follow Laplace distribution ${\rm Lap}(\nu^k)$ with $(\sigma_i^k)^2=2(\nu^k)^2$ satisfying Assumption \ref{assumption:dp-noise}, then:
\begin{enumerate}
\item Algorithm 1 is  $\epsilon$-differentially private with the cumulative privacy budget from $k=0$ to $k=T_0$ bounded by $\epsilon\leq \sum_{k=1}^{T_0}\frac{2\bar{C}\lambda^{k}}{\nu^{k}}$,  where $\bar{C}$ is from Assumption \ref{as:bounded_gradients}. And the cumulative privacy budget is always finite even as $T_0\rightarrow\infty$  when the sequence  $\{\frac{\lambda^{k}}{\nu^k}\}$ is summable;
\item If  non-negative sequences  $\{\nu'^k\}$ and $\{\lambda^k\}$ satisfy $\Phi_{\lambda,\nu'}\triangleq \sum_{k=1}^{\infty}\frac{\lambda^k}{\nu'^k}<\infty$, then picking  DP noise parameter as $\nu^k=\frac{2 \bar{C}\Phi_{\lambda,\nu'}}{ \epsilon  }\nu'^k$ ensures that Algorithm 1 is $\epsilon$-differentially private with any cumulative privacy budget $\epsilon>0$, even  when the   iteration number tends to infinity;
\item In the special case  $\lambda^k=\frac{1}{k}$ and $\gamma^k=\frac{1}{k^{0.9}}$,   setting $\nu^k=\frac{2\bar{C} \Phi}{\epsilon}k^{0.3}$ with  $\Phi\triangleq \sum_{k=1}^{\infty}\frac{1}{k^{1.3}}\approx 3.93$ (which  satisfies Assumption \ref{assumption:dp-noise}) ensures that  Algorithm 1 is  $\epsilon$-differentially private with any cumulative privacy budget $ \epsilon>0$ even when the  iteration number tends to infinity.
\end{enumerate}
\end{Theorem 1}
\begin{proof}

We first show that the sensitivity $\Delta^k$ of  Algorithm 1 is no larger than $ 2\bar{C}\lambda^{k-1}$. According to Definition \ref{de:sensitivity}, for any given  observation $\mathcal{O}$ and initial state $\vartheta^0$, the sensitivity at iteration $k$ is determined by $ \|\vartheta^{k+1}-\vartheta'^{k+1}\|_1$ where  $\vartheta^{k+1}\in\mathcal{R}_{\mathcal{P}, \vartheta^0}^{-1}(\mathcal{O})$ and $\vartheta'^{k+1}\in\mathcal{R}_{\mathcal{P'}, \vartheta^0}^{-1}(\mathcal{O}) $.
Since $\mathcal{P}$ and $\mathcal{P'}$ are adjacent, only one of their cost functions is different. Pick  this different cost function as the  $i$th one, i.e., $f_i(\cdot)$,  without loss of generality. Given that the observations under  $\mathcal{P}$ and $\mathcal{P'}$  are identical, we have
$
 x_{(i)\ell}^k={x}'^{k}_{(i)\ell}$ for all $k\geq 0$ and $\ell\neq i.
$

By defining $x_{(j):}^k\triangleq [(x_{(j)1}^k)^T,\,\cdots,\,(x_{(j)m}^k)^T]^T$, we have
\[
\begin{aligned}
&\|{ \vartheta^{k+1}-\vartheta'^{k+1}} \|_1= \\
&\left\| \left[ (x_{(1):}^{k+1})^T,\cdots,(x_{(m):}^{k+1})^T \right]^T -\left[({x'}_{(1):}^{k+1})^T, \cdots,({x'}_{(m):}^{k+1})^T \right]^T\right\|_1\\
&
=\left\| \left[\begin{array}{c}x_{(i):}^{k+1}-{x'}_{(i):}^{k+1} \end{array}\right] \right\|_1 =\left\| \left[\begin{array}{c}x_{(i)i}^{k+1}-{x'}_{(i)i}^{k+1} \end{array}\right] \right\|_1,
\end{aligned}
\]
where in the second-to-last  equality we used the fact that only the $i$th cost function is different, and in the last equality, we used the fact that $x_{(i)\ell}^{k+1}$ and ${x'}_{(i)\ell}^{k+1}$ for $\ell\neq i$ are updated independently of $F_i(\cdot,\cdot)$ and $F'_i(\cdot,\cdot)$, and hence are the same when observations are  identical in $\mathcal{P}$ and $\mathcal{P'}$.

Using
 (\ref{eq:update_in_Algorithm1}),  we can further write the above relationship as
\[
\begin{aligned}
&\|{ \vartheta^{k+1}-\vartheta'^{k+1}} \|_1=\\
&\hspace{-0.1cm}\left\|\hspace{-0.03cm}x_{\hspace{-0.03cm}(i)i}^{k}\hspace{-0.08cm}+\hspace{-0.08cm}\gamma^{k}\hspace{-0.10cm}\textstyle\sum_{j\in \mathbb{N}_i^{\rm in}} \hspace{-0.06cm}L_{ij}(x_{\hspace{-0.03cm}(j)i}^{k}\hspace{-0.08cm}+\hspace{-0.08cm}\zeta_{\hspace{-0.03cm}(j)i}^{k}\hspace{-0.08cm}-\hspace{-0.08cm}x_{\hspace{-0.03cm}(i)i}^{k})\hspace{-0.08cm}-\hspace{-0.08cm}\lambda^{k} \hspace{-0.05cm} F_i(x_{\hspace{-0.03cm}(i)i}^{k},x_{\hspace{-0.03cm}(i)-i}^{k})\right.\\
& \left.-\hspace{-0.03cm} {x'}_{\hspace{-0.1cm}(i)i}^{k} \hspace{-0.08cm} - \hspace{-0.08 cm}\gamma^{k}\hspace{-0.1cm}\textstyle\sum_{j\in \mathbb{N}_i^{\rm in}} \hspace{-0.1cm} L_{ij}({x'}_{\hspace{-0.12cm}(j)i}^{k} \hspace{-0.1cm}+\hspace{-0.1cm}{\zeta'}_{\hspace{-0.12cm}(j)i}^{k}\hspace{-0.1cm}-\hspace{-0.1cm}{x'}_{\hspace{-0.12cm}(i)i}^{k}) \hspace{-0.1cm}+\hspace{-0.1cm}\lambda^{k}\hspace{-0.08cm} F'_i({x'}_{\hspace{-0.12cm}(i)i}^{k},{x'}_{\hspace{-0.12cm}(i)-i}^{k})\right \|_1 \\
& \leq\left\|  \lambda^{k} F_i(x_{(i)i}^{k},x_{(i)-i}^{k})  -   \lambda^{k} {F'}_i({x'}_{(i)i}^{k},{x'}_{(i)-i}^{k}) \right\|_1,
\end{aligned}
\]
where we have used  the fact that the shared messages are the same for iterations up to $k$.
According to Assumption \ref{as:bounded_gradients}, we have
$
\|F_i(x_{(i)i}^{k},x_{(i)-i}^{k})\|_1\leq \bar{C}$, and $\|{F'}_i({x'}_{(i)i}^{k},{x'}_{(i)-i}^{k})\|_1\leq \bar{C}.
$
Combining the two preceding relations leads to
$
\|{ \vartheta^{k+1}-\vartheta'^{k+1}} \|_1
 \leq 2\lambda^{k}\bar{C}.
$

Using Lemma \ref{Le:Laplacian},  the cumulative privacy budget is always less than $ \sum_{k=1}^{T_0}\frac{2\bar{C}\lambda^k}{\nu^k}$. Therefore, the cumulative privacy budget $\epsilon$ will always be finite even when the number of iterations $T_0$ tends to infinity if  $\{\frac{\lambda^k}{\nu^k}\}$ satisfies $\sum_{k=0}^{\infty}\frac{\lambda^k}{\nu^k}<\infty$, which concludes the proof for the first statement.

The preceding derivation shows that the cumulative privacy budget is inversely proportional to $\nu^k$. Thus, the second statement can be obtained from the first statement by  scaling $\nu^k$. The   third statement follows by specializing the selection of $\lambda^k$, $\gamma^k$, and $\nu^k$ sequences as specified in the statement.
\end{proof}

It is worth noting that to ensure a finite cumulative  privacy budget,  \cite{ye2021differentially} and \cite{Huang15}  resort to  summable stepsizes, which, however, also make provable  convergence to the exact desired value impossible. To the contrary,  our approach  allows the stepsize sequence to be non-summable, and hence ensures  both provable convergence and finite cumulative privacy budget, even when the number of iterations goes to infinity.   This appears to be  the first time that both { provable} convergence and rigorous $\epsilon$-differential privacy   is achieved in general networked games on directed communication graphs.

\section{Numerical Simulations}\label{se:simulation}
We use the  networked Nash-Cournot game in \cite{pavel2019distributed,koshal2016distributed,nguyen2022distributed} to evaluate our  approach. Due to space limitations, we suppress  the application details of this game and only provide  the mathematical representation. More specifically, we consider 20 players with each player having a cost function
$
 f_i(x_i,x)=x_i^TQ_ix_i+q_i^Tx_i-(\bar{P}-\Xi Bx)^T B_i^T x_i$, where $x_{i}\in\mathbb{R}^{d_i}$ with $1\leq d_i\leq N$.  $Q_i\in\mathbb{R}^{d_i\times d_i}$ is a randomly generated positive definite matrix and $q_i\in\mathbb{R}^{d_i}$. $\bar{P}$ is a positive vector and  $\Xi$ is a diagonal matrix with positive diagonal entries, both of which are randomly chosen in the numerical simulation. $B$ is constructed as $B\triangleq [B_1,\,\cdots,B_N]$, where $B_i\in\mathbb{R}^{N\times d_i}$ is chosen following \cite{nguyen2022distributed}.  The   communication graph  is  generated   randomly but assuring that it is  strongly connected.  

\begin{figure}
\center
\includegraphics[width=0.35\textwidth]{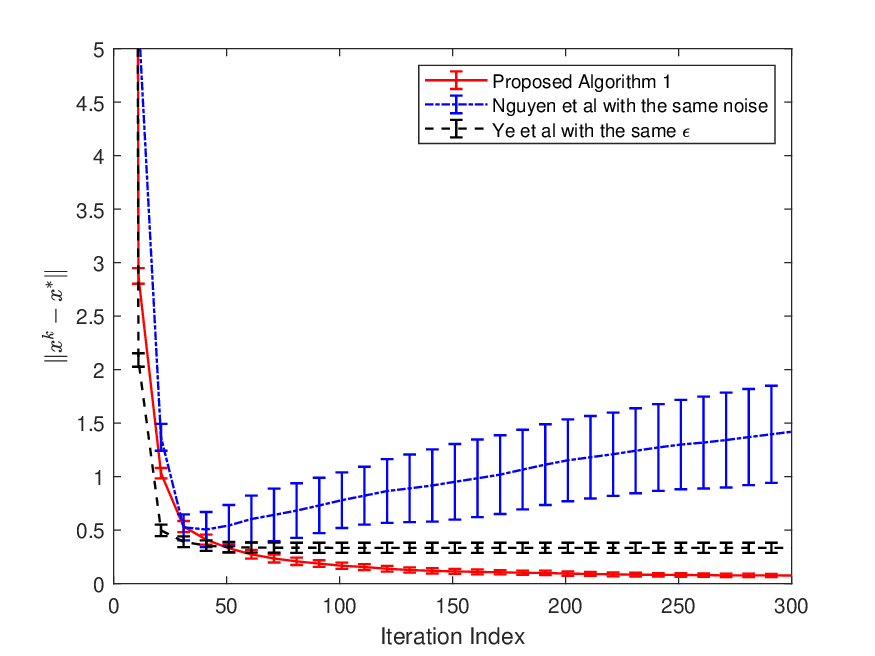}
    \caption{Comparison   with the existing distributed NE seeking algorithm  in \cite{nguyen2022distributed} (under the same noise level) and the  differential-privacy approach for aggregative games in \cite{ye2021differentially}  (under the same privacy budget $\epsilon$).}
    \label{fig:comparison_algo1}
    \vspace{-0.2cm}
\end{figure}

  To evaluate  the proposed approach,  we  inject    vector noise $\zeta_{(i)\ell}^k$ ($1\leq \ell\leq 20$) in every message $x_{(i)\ell}^k$ that  player $i$  shares  in every iteration. Each element of the noise vector $\zeta_{(i)\ell}^k$ follows Laplace  distribution with parameter $\nu^k=1+0.1k^{0.2}$.   We set  $\lambda^k=\frac{0.1}{1+0.1k}$ and $\gamma^k=\frac{1}{1+0.1k^{0.9}}$, respectively, which  satisfy the conditions in Theorems 1 and  2. We ran our Algorithm 1 for 100 times and calculated the average of the gap $\|x^k-x^{\ast}\|$ between the generated iterate $x^k$ and the NE $x^{\ast}$  as a function of  $k$. We also calculated the variance of the gap of the 100 runs as a function of  $k$. The trajectories of the average and variance are  given by the red curve and error bars in Fig. \ref{fig:comparison_algo1}. For comparison, we also ran the  distributed NE seeking algorithm  proposed by Nguyen et al. in \cite{nguyen2022distributed} under the same noise level, and the DP approach for networked games proposed by Ye et al. in \cite{ye2021differentially} under the same  privacy budget $\epsilon$. Note that  \cite{ye2021differentially} addresses undirected graphs but its  DP strategy, i.e.,  geometrically decreasing stepsizes for a finite privacy budget, can be adapted to the directed-graph scenario.   The  average errors/variances of the two approaches are given by  the blue and black curve/error bars in Fig. \ref{fig:comparison_algo1}.  The comparison clears shows that our approach has a much better  accuracy.

\section{Conclusions}\label{se:conclusions}
This paper has introduced  a  distributed NE seeking approach that can ensure  both { provable} convergence and rigorous $\epsilon$-DP,  even when the number of iterations tends to infinity. The simultaneous achievement of both goals is in sharp contrast to existing DP solutions for  aggregative games that trade provable convergence for privacy, and to  our knowledge, has not been achieved before for general networked Nash games. The approach is applicable to general directed graphs. Numerical   results   confirm effectiveness of the proposed approach.


\bibliographystyle{IEEEtran}

\bibliography{reference1}

\end{document}